\documentclass[twocolumn,aps,groupedaddress]{revtex4}

\usepackage{graphicx}
\usepackage{dcolumn}
\usepackage{bm}
\usepackage{amsmath,amssymb,amsfonts,esint}
\usepackage{color}
\usepackage{ulem}
\usepackage{multirow}
\usepackage[caption=false]{subfig}

\begin{document}

\title{Classification of Critical Points in Energy Bands Based on \\ Topology, Scaling and Symmetry}
\author{Noah F. Q. Yuan}
\author{Liang Fu}
\affiliation{Department of Physics, Massachusetts Institute of Technology, Cambridge, Massachusetts 02139, USA}

\begin{abstract}
A critical point of the energy dispersion is the momentum where electron velocity vanishes. At the corresponding energy, the density of states (DOS) exhibits non-analyticity such as divergence.
Critical points can be first classified as ordinary and high-order ones, and the ordinary critical points have been studied thoroughly by L\'eon van Hove. In this work, we describe and classify high-order critical points based on topology, scaling and symmetry, which are beyond L\'eon van Hove's framework. %Among them, topology describes asymptotical behavior far from critical points, while scaling and symmetry describe asymptotical behavior close to critical points. 
We show that high-order critical points can have power-law divergent DOS with particle-hole asymmetry, and can be realized at generic or symmetric momenta by tuning a few parameters such as twist angle, strain, pressure and/or external fields.
\end{abstract}

\maketitle

The density of states (DOS) of a solid plays an important role in the thermodynamic properties. A large DOS can enhance response functions such as optical absorption and heat capacity at single-particle level, and magnify various electronic instabilities such as magnetism and superconductivity with interactions. DOS can also affect transport properties significantly.

The DOS in energy domain is determined by band dispersion in the Brillouin zone (BZ), and large DOS is usually related to the so-called {\it critical point} in BZ, where electron velocity vanishes and DOS exhibits non-analyticity such as divergence at the corresponding energy.
Since electron velocity is zero, the expansion of energy dispersion near a critical point will be at least second order in momentum deviation. When all the second order terms are nonzero, we call this critical point an ordinary one, otherwise a high-order critical point.

L\'eon van Hove studied the ordinary critical points in the context of phonons \cite{VHS}, where he found in two dimensions (2D), due to the nontrivial topology of BZ, {every} smooth energy dispersion will have at least two saddle points if critical points are all ordinary. At the corresponding energy, DOS is logarithmically divergent and is now known as the van Hove singularity (VHS).

The high-order critical points have been proposed occasionally in specific materials under different names in the context of DOS singularity, including the so-called extended VHS \cite{EVHS1,EVHS2cu,EVHS3,EVHS4,EVHS5cu,EVHS6cu,EVHS7}, multicritical points \cite{tricritical1,tricritical2,Sr2Ru3O7,Sr2Ru3O7a} and high-order VHS \cite{magic}. At the energy of a high-order critical point, the DOS can be power-law divergent, stronger than ordinary VHS, and hence we call it a high-order VHS of DOS.

In this work, we provide systematic analysis of high-order critical points. We mainly focus on two dimensions (2D), but also include 1D and 3D cases for completeness. Our analysis is based on three general principles: Topology, scaling and symmetry.

Topology of a critical point describes the asymptotic behavior away from the point, which has been applied by L\'eon van Hove in studying ordinary critical points. When critical points are not limited to ordinary ones, topology still applies and the general theory includes L\'eon van Hove's results as a corollary. 

Scaling and symmetry of a critical point describe the asymptotic behavior close to the point, which are beyond L\'eon van Hove's framework. As we approach a critical point in momentum space, the energy also approaches the DOS singularity energy, described by a scaling law specific to the critical point. In terms of DOS, the scaling property of a critical point is reflected not only in the divergence (i.e. logarithmic versus power-law), but also the particle-hole symmetry: DOS of ordinary critical points is particle-hole symmetric, while that of high-order critical points can be particle-hole asymmetric. 
%The particle-hole asymmetry of high-order VHS can become important when interactions are considered.

We find that high-order critical points can be generally realized by tuning the system with one or more parameter such as twist angle, strain, pressure and/or external fields. When the critical point is at a generic point without additional symmetries, the number of tuning parameters is in fact determined by scaling too. The number of tuning parameters can be reduced when the high-order critical point is at a symmetric point. 

Recent experimental progress in 2D moir\'e superlattices such as twisted bilayer graphene \cite{TBG1,TBG2,TBG3,Kerelsky} and graphene heterostructures \cite{TLG1,TLG2,TLG3,Constantine} has enabled us to continuously tune electronic band structures in 2D systems, where high-order VHS can be realized at symmetric points \cite{magic} or generic positions \cite{Strain}, and could be relevant to experimentally observed insulating and superconducting phases at magic angle \cite{TBG1,TBG2,TBG3}.

This work is about the single-particle physics at zero temperature. Interaction effects between ordinary saddle points have been studied intensively \cite{{hirsch},{Dzya},{schulz},{lederer},{furukawa},{hur},{raghu},{nandkishore},{gonz},{isobe}}, and some theoretical treatments have been carried out for high-order critical points \cite{tricritical1,supermetal}. The high-order critical points can also affect transport properties at finite temperature such as thermoelectric effect \cite{mahan}.

This paper is structured as follows. In Sec. \ref{os}, we define ordinary and high-order critical points \cite{VHS,magic}. We then describe the topology, scaling and symmetry of (high-order) critical points in 1D (Sec. \ref{1d}), 2D (Sec. \ref{gs}) and 3D (Sec. \ref{g3d}) respectively.
Finally in Sec. \ref{per} we focus on high-order critical points that can be realized by tuning one parameter.
Our results also apply to other quasiparticles such as phonons and magnons.

\section{Ordinary and High-Order Critical Points}\label{os}
A critical point is the momentum where electron velocity
vanishes, which can be local extrema or saddle points of the energy
dispersion. Throughout this paper we
set the critical point to be at zero energy $E = 0$ and
zero momentum $\bm p =\bm 0$ if not specified
otherwise.

One can expand the energy dispersion in the vicinity of $\bm p=\bm 0$. To the second order, the Taylor expansion is
\begin{eqnarray}\label{eq_os}
E=D_{ij}p_{i}p_{j},
\end{eqnarray}
where $  D $ is the symmetric Hessian matrix. Eq. (\ref{eq_os}) is homogeneous and second order, satisfying $ E(\lambda^{\frac{1}{2}}p_x,\lambda^{\frac{1}{2}}p_y)=\lambda E(p_x,p_y) $ with arbitrary $ \lambda>0 $.
By a linear coordinate transform $ \tilde{\bm p}=U\bm p $ ($U$ is a rotation matrix in SO(2)), the dispersion can be written as the canonical form in the new coordinate
\begin{eqnarray}\label{eq_osc}
E=\alpha\tilde{p}_{x}^2+\beta\tilde{p}_{y}^2,
\end{eqnarray}
where $ \alpha,\beta\neq 0 $ are eigenvalues of $ D $.
When $ \alpha,\beta>0 $ the critical point is a local minimum, when $ \alpha,\beta<0 $ a local maximum, and when $ \alpha\beta<0 $ a saddle point.% also known as van Hove singularity (VHS).

At a given energy, the energy contour near an extremum (minimum or maximum) is a closed loop, and the DOS near an extremum has a discontinuity at $ E=0 $. The energy contour near an ordinary saddle point consists of two curves, which intersects at $ \bm p=\bm 0 $ when $ E=0 $. The DOS near such saddle point is logarithmically divergent \cite{VHS}.% known as van Hove singularity (VHS).

\begin{figure}
\includegraphics[width=0.5\textwidth]{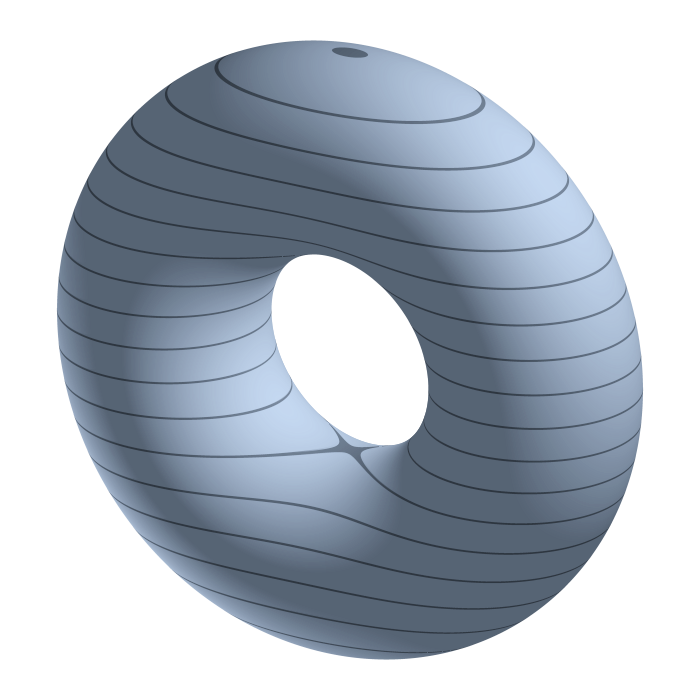}
\centering
\caption{Energy contours of a smooth energy dispersion defined on the torus.}\label{fig_0}
\end{figure}

Since $ \alpha\beta\neq 0 $, the critical point is called {\it ordinary}, which is a {\it nondegenerate} quadratic form of the momentum $(p_x, p_y)$.
When all critical points are ordinary (\ref{eq_os}), due to the toric topology of 2D Brilluoin zone, there exist at least one minimum, one maximum and two saddle points \cite{VHS}. To illustrate this, one can draw energy contours on the torus as shown in Fig. \ref{fig_0}. Since the torus is finite, the contours will start and end with points, which correspond to a minimum and a maximum. Furthermore, there are two types of energy contours on the torus, i.e. the contractible and non-contractible loops. By continuity, there will exist two transition points where a contractible loop bifurcates into two non-contractible loops, which correspond to two ordinary saddle points.

However, the topological argument above in fact only guarantees the existence of critical momentum points, but does not specify the detailed dispersion in their vicinity. When the energy dispersion is tuned continuously with one or more external parameters (such as pressure, strain, displacement field, magnetic field etc), the quadratic form of energy dispersion near a critical point can be made degenerate (i.e. $\det D=0$), and then high-order terms must be included in the Taylor expansion. The resulting {\it high-order} critical point is the subject of this study.

As a concrete example, we consider the high-order critical point introduced in Ref. \cite{magic}, encountered when one of the quadratic coefficients is tuned to zero, 
\begin{eqnarray}\label{eq_hs}
E=\alpha p_{x}^2+\gamma p_{x}p_{y}^2+\kappa p_{y}^4.
\end{eqnarray}
This dispersion is quasihomogeneous with scaling property $ E(\lambda^{\frac{1}{2}}p_x,\lambda^{\frac{1}{4}}p_y)=\lambda E(p_x,p_y) $. Under the nonlinear coordinate transform
$ \tilde{p}_x=p_{x},\tilde{p}_y=p_{y}+\gamma p_{x}^2/(2\alpha) $, the dispersion can be recast into the canonical form
\begin{eqnarray}\label{eq_hsc}
E=\alpha\tilde{p}_{x}^2+\left(\kappa-\frac{\gamma^2}{4\alpha}\right)\tilde{p}_{y}^4.
\end{eqnarray}
When $ 4\alpha\kappa -\gamma^2>0 $ the critical point is a local minimum ($ \alpha>0 $) or maximum ($ \alpha<0 $), and when $ 4\alpha\kappa -\gamma^2<0 $ a saddle point. The energy contour near an extremum is still a closed loop, and the energy contour near a saddle point is still formed by two curves. When $4\alpha\kappa -\gamma^2=0$, even higher order terms need to be included.

Unlike ordinary saddle points, the two energy contours at energy $ E=0 $ touch \textit{tangentially} at the high-order saddle point $\bm p=\bm 0$, and the DOS is power-law divergent with exponent $-\frac{1}{4}$ \cite{magic}. %As will be discussed, DOS near high-order extrema is also power-law divergent with the same exponent.

In addition to the difference in DOS divergence and scaling property of energy dispersion, ordinary and high-order critical points also exhibit different behaviors under perturbations. Perturbations could generate new terms in the Taylor expansion besides modifying the original dispersion.
An ordinary critical point is robust under perturbations, as zeroth order $E_0$ and linear perturbation $ \bm V\cdot\bm p $ only shift the momentum and energy of critical point, second order perturbations will slightly modify the Hessian matrix, and third or higher order perturbations do not affect the asymptotic behavior near $\bm p=\bm 0$.
In contrast, the high-order saddle point in (\ref{eq_hs}) can split into two ordinary saddle points and one local extrema under perturbations \cite{magic}.
As will be shown later, the DOS divergence and response to perturbations are both closely related to the scaling behavior of energy dispersion.

Besides (\ref{eq_hs}), there exists many other types of high-order critical points, some of which requires tuning more than one parameter. 
We would like to classify different types of (high-order) critical points within an isolated band and determine the appropriate Taylor expansion of energy dispersion in the vicinity of critical point $\bm p=\bm 0$:  
\begin{eqnarray}\label{eq_taylor}
\mathcal{E}=D_{ij}p_{i}p_{j}+\Gamma_{ijk}p_{i}p_{j}p_{k}+\dots.
\end{eqnarray}
From the examples of ordinary and high-order critical points discussed in this Section, we find that scaling property is important in determining the allowed form of dispersions near critical points. In addition, the topology of the critical point (i.e. saddle point or local extrema) constrains the sign of coefficients in the Taylor expansion.

\section{High-Order Critical Points in 1D}\label{1d}
As discussed above, our strategy is to use topology and scaling to classify critical points. To illustrate our strategy explicitly, we analyze critical points in 1D in terms of topology and scaling. The dispersion can be easily written down as $\mathcal{E}=D_n p^n$ with $n=2$ denoting ordinary and $n\geqslant 3$ denoting high-order, and we then need to determine integer $n$ and coefficient $D_n$ by topology and scaling.

In 1D, the carrier is either a left-mover or a right-mover. On the two sides away from a critical point $p=0$, carriers can have the same or opposite types of movers, which is described by the {\it topological index}
\begin{eqnarray}\label{eq_pi1D}
I\equiv\frac{1}{2}[{\rm sgn}v(+\delta)-{\rm sgn}v(-\delta)],
\end{eqnarray}
where $ v=d\mathcal{E}/dp $ is the velocity and $p=0$ is the only critical point in the interval $[-\delta,+\delta]$.
For critical point $\mathcal{E}=D_n p^n$, the topological index is determined by the sign of coefficient
$
I =\frac{1}{2}[1+(-1)^n]{\rm sgn}(D_n).
$ For example, ordinary critical point has topological index $I={\rm sgn}(D_2)=\pm$ while a saddle point $n=3$ has topological index $I=0$. In this way, we relate both integer $n$ and coefficient $D_n$ to topology.

As a byproduct, since our dispersion is defined with periodic boundary condition, after winding around the BZ (a circle), the velocity should go back to itself. As a result, the total topological index over the whole BZ should be zero
\begin{eqnarray}\label{eq_p1d}
\sum_i I_i=0.
\end{eqnarray}
Since dispersion is bounded, it will at least have a minimum and a maximum, which already satisfies the topological constraint (\ref{eq_p1d}). Hence there can be no saddle point in 1D. The results can be different in 2D and 3D as discussed in later Sections.

Next we find the scaling property $\mathcal{E}(\lambda p)=D_n (\lambda p)^n=\lambda^n E(p)$ is purely determined by the integer $n$. There are two ways to reveal the scaling property. First, one can compute the DOS $\rho$, which diverges $\rho(\mathcal{E})\propto |\mathcal{E}|^{\nu}$ with power law exponent $\nu =\frac{1}{n}-1$, for example at an ordinary critical point $n=2,\nu=-\frac{1}{2}$. 

Second, we can add perturbations $h_j$ to the dispersion $\mathcal{E}=D_np^n+\sum_{j=1}^{n-1}h_j p^j$. As $\mathcal{E}$ is a polynomial of degree $n$, $d\mathcal{E}/dp$ is of degree $n-1$ and the equation $d\mathcal{E}/dp=0$ has at most $n-1$ real solutions. In other words, the critical point $E(p)=D_n p^n$ will split into at most $\mu =n-1$ critical points under perturbations.

In above discussions, we derive topology and scaling from the explicit form of energy dispersion. However, we can also do it reversely, i.e. derive the explicit form of energy dispersion from topology and scaling. This can be written as the one-to-one correspondence $ (n,{\rm sgn}D_n)\leftrightarrow (I,\nu)\leftrightarrow (I,\mu) $, provided the {\it compatibility conditions} $I=\mu$(mod 2) and $I^2\leqslant 1$. Since $|D_n|$ can not be reproduced, the dispersion obtained in this strategy is unique up to an invertible and smooth coordinate transform.

We will employ this strategy and generalize results in this Section to 2D and 3D in Sec. \ref{gs} and \ref{g3d} respectively, where symmetry also constrains the form of energy dispersion.

\section{High-Order Critical Points in 2D}\label{gs}
In this Section, we classify critical points and hence DOS singularities of 2D Taylor expansion (\ref{eq_taylor}) according to topology, scaling properties and symmetry.

\subsection{Topology}
%Similar to 1D, in 2D a \textit{critical} momentum point at $ \bm p=\bm p_0 $ of dispersion $\mathcal{E}(\bm p)$ is formally defined by the following condition
%\begin{eqnarray}
%\nabla_{\bm p}\mathcal{E}|_{\bm p=\bm p_0}=\bm 0. 
%\end{eqnarray}
We now introduce an integer topological index for the critical point $\bm p=\bm 0$ in 2D
\begin{eqnarray}\label{eq_pi}
I\equiv\frac{1}{2\pi}{\rm Im}\oint_{\mathcal{C}}\frac{dv_{x}+idv_{y}}{v_{x}+iv_{y}}, 
\end{eqnarray} 
which is the winding number of the velocity field $ \bm v=\nabla_{\bm p}\mathcal{E} $ around the critical point $\bm 0 $. Here the counterclockwise contour $\mathcal{C}$ encloses a single critical point $ \bm 0 $ only. To ensure that $\mathcal{C}$ always exists, it is assumed that the velocity field only vanishes at {\it isolated}  points. This condition is expected to hold generically, as the alternative scenario of a line of critical points generally requires tuning an infinite number of control parameters.

As concrete examples, we can work out the velocity $ \bm v=2 D\bm p $ of dispersion (\ref{eq_os}), and critical point $\bm p=\bm 0$ has topological index $ I={\rm sgn}(\det D)=\pm 1$ for extrema (+) or saddle points ($-$).

For the high-order dispersion (\ref{eq_hs}), the velocity is $ \bm v=(2\alpha p_x+\gamma p_y^2,2\gamma p_xp_y+4\kappa p_y^3) $ and the topological index of critical point $\bm p=\bm 0$ is $ I={\rm sgn}(4\alpha\kappa -\gamma^2) $. 

In both cases, the topological index is invariant under the coordinate transform of momentum (linear or nonlinear), as long as it is invertible and smooth. %In the following we will set the critical point to be at $ \bm p=\bm 0 $. 
In fact, the topological index of an isolated critical point can be computed by the following formula
\begin{eqnarray}\label{eq_top}
I=1-\left\lfloor\frac{n_{e}+n_{h}}{2}\right\rfloor
\end{eqnarray}
where $ n_e (n_h) $ is the number of segments in the energy contour at $E>0$ ($E<0$) within the region enclosed by $\mathcal{C}$, and $ \lfloor x\rfloor $ denotes the largest integer smaller than or equal to $ x $. 
From its definition (\ref{eq_pi}), we find the topological index is invariant under energy inversion $ E\to -E $, which is reflected by the exchange symmetry $ n_e\leftrightarrow n_h$ of Eq. (\ref{eq_top}).

\begin{figure}
\includegraphics[width=0.5\textwidth]{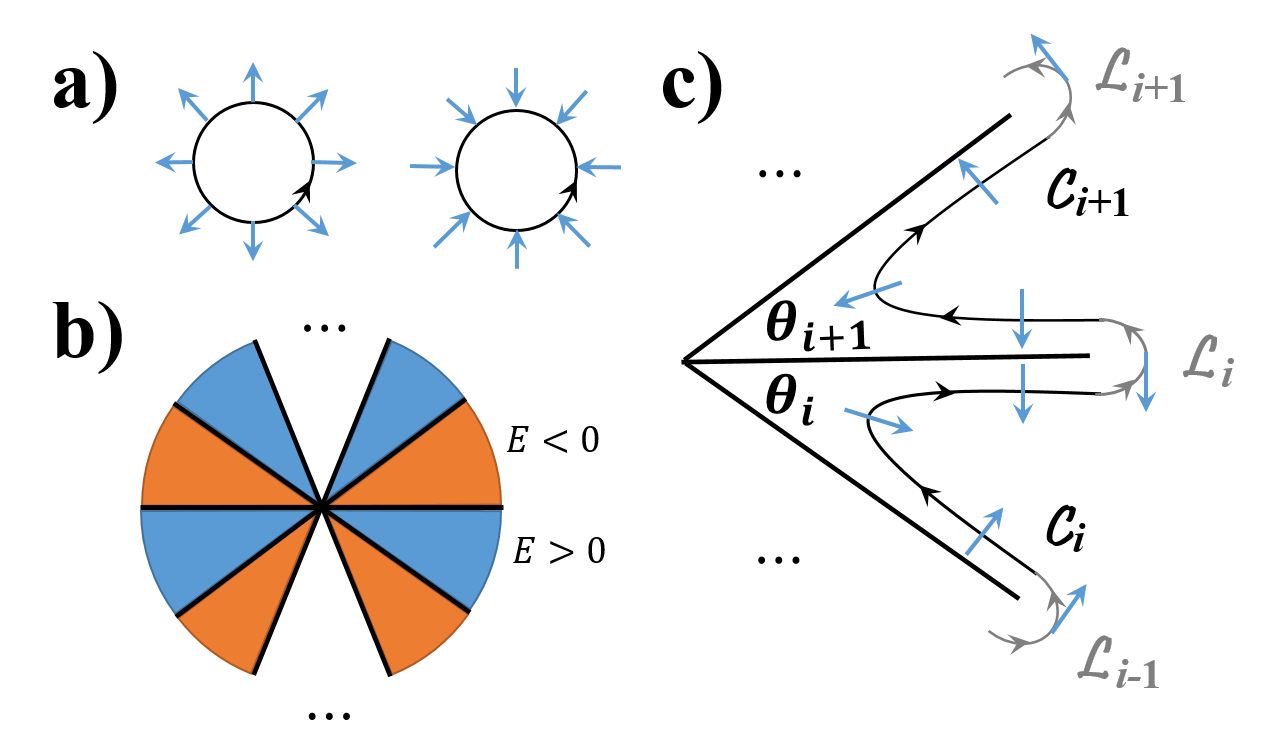}
\centering
\caption{{Winding of velocity vector (blue arrows) along the counterclockwise contour $ \mathcal{C} $ (solid arrowed curves). In a) local minimum (left) and local maximum (right), $\mathcal{C}$ is the energy contour at $E>0$ or $E<0$ respectively. In b) and c) saddle point, solid thick lines are energy contours at $E=0$ which divide the 2D plane into $2n$ sectors, and the gray line $\mathcal{L}_i$ connects energy contours $\mathcal{C}_i$ at $E>0$ in sector $i$ and $\mathcal{C}_{i+1}$ at $E<0$ in sector $i+1$, so that $ \mathcal{C}=\cap_{i=1}^{2n}(\mathcal{C}_{i}\cap\mathcal{L}_i) $. In a), b) and c) the velocity vector is perpendicular to the energy contour.}}\label{fig_1}
\end{figure}

To prove this formula, we just enumerate all cases of isolated critical points.
Near a local extremum, all the energy contours are closed loops centered around the local extremum itself, and we can then choose $\mathcal{C}$ to be an energy contour at $E\neq 0$. Since velocity vectors are perpendicular to energy contours, the counterclockwise winding angle of the velocity along $ \mathcal{C} $ is $ 2\pi $ (Fig. \ref{fig_1}a), and the topological index of a local extremum is $I=1$. %On the other hand, we find $ n_e=1,n_h=0 $ for a local minimum, while $ n_e=0,n_h=1 $ for a local maximum. 
Formula (\ref{eq_top}) is thus valid for local extrema where $ n_e=1,n_h=0 $ for minimum, and $ n_e=0,n_h=1 $ for maximum.

On the contrary, near a saddle point, the energy contours are not contractible. Instead, the energy contour at $E= 0$ consists of several curves intersecting at the critical point, which divide the 2D momentum space into different sectors.
%If the energy contour at $E>0$ has $n_{e}=n\geqslant 1$ segments in separate momentum sectors, by continuity of the energy dispersion in two dimensions, the energy contour at $E<0$ will also have $n_{h}=n$ segments in the complementary momentum sectors, 
By continuity of the energy dispersion in two dimensions, energy contours at $E>0$ and $E<0$ will form an alternating pattern in different momentum sectors, as shown in Fig. \ref{fig_1}b. Hence $n_e=n_h\equiv n$ and there are in total $2n$ sectors labeled by $ i=1,\dots 2n $, such that the sign of energy dispersion in sector $i$ is $ (-)^i $. We then choose $\mathcal{C}$ to be formed by energy contours $\mathcal{C}_{i}$ in all sectors, connected by lines $\mathcal{L}_i$ near sector boundaries, as shown in Fig. \ref{fig_1}c. Within each sector, the winding angle of velocity vector along $\mathcal{C}$ is $ \theta_{i}-\pi $, where $\theta_{i}$ is the angle of this sector. The total winding angle along $ \mathcal{C} $ hence reads $ \sum_{i=1}^{2n}(\theta_i -\pi)=\sum_{i=1}^{2n}\theta_i-2n\pi=2\pi(1-n) $, and the topological index of a saddle point is thus a non-positive integer given by $I=1-n\leqslant 0$. Formula (\ref{eq_top}) hence also applies to saddle points.

%When $\mathcal{C}$ encloses more than one isolated critical point, the calculated topological index will be the sum of topological indices of all enclosed critical points.
The total topological index of all critical points in the whole momentum space reflects the topology of the toric BZ, which is equal to its Euler characteristic \cite{topology}
\begin{eqnarray}\label{eq_charge}
\sum_{i} I_{i}=0.
\end{eqnarray}
This result is known as the Poincar\'e-Hopf theorem, which in fact applies to any dimension as shown in Sec. \ref{1d} and will be shown in Sec. \ref{g3d}.

A smooth and bounded energy dispersion will have at least one minimum and one maximum. According to Eq. (\ref{eq_charge}) there will also be at least one saddle point. %Hence, every continuous energy dispersion defined in the 2D BZ will have at least one minimum, one maximum and one saddle point with corresponding DOS singularities.
When all critical points are ordinary, there will be at least two ordinary saddle points, which corresponds to the case of L\'eon van Hove \cite{VHS}. When there is only one saddle point, according to our topological argument, its topological index must be $-2$, with three segments in each energy contour, which will be discussed in Secs. \ref{sp} and Appendix \ref{mod}.

\subsection{Scaling}

We rewrite the dispersion (\ref{eq_taylor}) as the sum of two parts
\begin{eqnarray}\label{eq_cata}
\mathcal{E}({\bm p})=E(\bm p)+\bm h\cdot\bm H(\bm p).
\end{eqnarray}
The first part, \textit{canonical} dispersion $E(\bm p)$ has vanishing linear terms $ \nabla_{\bm p}E|_{\bm p=\bm 0}=\bm 0 $ and is scale invariant
\begin{eqnarray}\label{eq_QH}
E(\lambda^{a} p_{x},\lambda^{b}p_{y})=\lambda E(p_{x},p_{y}),\quad (\lambda>0)
\end{eqnarray}
with scaling exponents $ a,b>0 $.  
All possible scaling exponents of the analytic energy dispersions will be worked out in the next subsection.

The second part, perturbation $\bm H$ is formed by monomials not satisfying scaling property (\ref{eq_QH}) and $ \bm h $ denotes corresponding Taylor coefficients. For every monomial perturbation $ H(\bm p) $ (a component of $\bm H$), we can work out its scaling behavior under scaling transform in (\ref{eq_QH})
\begin{eqnarray}
H(\lambda^{a} p_{x},\lambda^{b}p_{y})=\lambda^{\gamma} H(p_{x},p_{y}),\quad \gamma\neq 1.
\end{eqnarray}
As we approach the critical point $ \lambda\to 0^{+} $, the perturbation $H$ becomes relevant when $\gamma<1$ and irrelevant when $ \gamma> 1 $.

We have separated canonical dispersion and perturbations in (\ref{eq_taylor}) according to scaling property. In the following two subsections we then study canonical dispersion and perturbations respectively.

\subsection{Canonical dispersion}
Besides scaling property, analyticity and topology will impose constraints to the canonical dispersion. In this subsection we show that the interplay between scaling property, analyticity and topology will eventually determine the possible values of scaling exponents and the corresponding canonical dispersions.

Since canonical dispersion is analytic, we can write
\begin{eqnarray}\label{eq_can}
E(\bm p)=\sum_{m+n\geqslant 2}c_{mn}p_{x}^mp_{y}^n\quad (m,n=0,1,2,\dots),
\end{eqnarray}
where vanishing of linear terms leads to the condition $ m+n\geqslant 2 $. Here we construct a one-to-one mapping between the monomial $ p_{x}^mp_{y}^n $ and the ordered integer pair $(m,n)$.
Combining the analyticity condition (\ref{eq_can}) and scaling property (\ref{eq_QH}), we find the scaling exponents $a,b$ should satisfy $ ma+nb=1 $ if monomial $ (m,n) $ is allowed in the canonical dispersion. This motivates us to construct canonical dispersion as follows.

\begin{figure}
\includegraphics[width=0.4\textwidth]{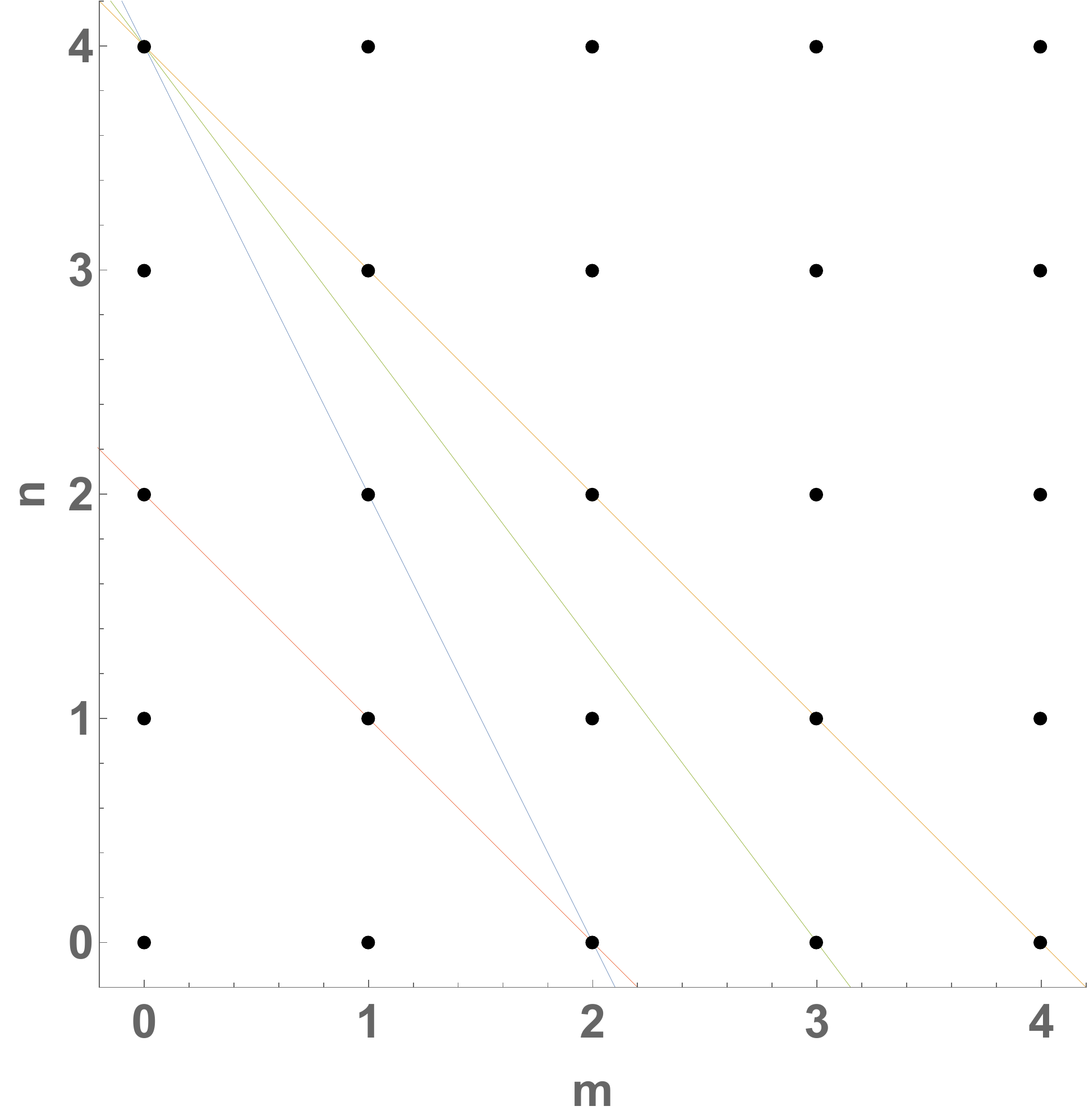}
\centering
\caption{{Integer grid in 2D, where each dot denotes a pair of integers $(m,n)$ with $ 0\leqslant m,n\leqslant 4 $, and each line $ ma+nb=1 $ denotes a pair of scaling exponents $ (a,b) $ with $ a,b>0 $. There are two pairs of integers on the green line, 3 pairs of integers on the red and blue lines and 5 pairs of integers on the yellow line.}}\label{fig_mn}
\end{figure}

We first choose two pairs of integers $ (m_1,n_1) $ and $(m_2,n_2)$
to determine scaling exponents $ a=(n_2-n_1)/\Delta,b=(m_1-m_2)/\Delta $, where
$ \Delta\equiv m_1n_2-m_2n_1\neq 0 $ and $ (m_1-m_2)(n_1-n_2)<0 $. To make $ \bm p=\bm 0 $ a critical point, we further require $ m_i+n_i\geqslant 2 $ for $ i=1,2 $.
After determining scaling exponents $a,b$, we then work out all integer solutions $(m,n)$ to the equation $ma+nb=1$ and hence determine the form of canonical dispersion. To find out all integer solutions, we can draw the line $ma+nb=1$ in the 2D integer grid as shown in Fig. \ref{fig_mn}, and count all integer dots on the line.
In the following we consider some specific examples to work out this procedure explicitly.

When we choose integer pairs $ (2,0) $ and $(0,2)$ we find $a=\frac{1}{2},b=\frac{1}{2}$, and there are three pairs of integer solutions to $ \frac{1}{2}m+\frac{1}{2}n=1 $ as shown in the red line of Fig. \ref{fig_mn}, which correspond to three terms $ p_{x}^2,p_{x}p_{y},p_{y}^2 $ in the dispersion (\ref{eq_os}) of an ordinary critical point.

When we choose integer pairs $ (2,0) $ and $(0,4)$ we find $a=\frac{1}{2},b=\frac{1}{4}$, and there are three pairs of integer solutions to $ \frac{1}{2}m+\frac{1}{4}n=1 $ as shown in the blue line of Fig. \ref{fig_mn}, which correspond to three terms $ p_{x}^2,p_{x}p_{y}^2,p_{y}^4 $ in the dispersion (\ref{eq_hs}) of a high-order critical point.

When we choose integer pairs $ (3,0) $ and $(0,4)$ we find $a=\frac{1}{3},b=\frac{1}{4}$, and there are only two pairs of integer solutions to $ \frac{1}{3}m+\frac{1}{4}n=1 $ as shown in the green line of Fig. \ref{fig_mn}, which correspond to the high-order critical point $E_6$ discussed in subsection \ref{sp}.

In fact the two pairs of integers cannot be chosen arbitrarily. For example, $ma+nb=1$ cannot be parallel to one of the axes. Formally, this is a topological requirement: We require the canonical dispersion to contain isolated critical points only.
As a result, the canonical dispersion has to contain at least two terms, which are either $ \{p_{x}^m, p_{y}^n\} $, $ \{p_{x}^{m} p_{y}, p_{y}^n\} $, $ \{p_{x}^{m}, p_{x}p_{y}^n\} $ or $ \{ p_{x}^{m}p_y, p_{x}p_{y}^{n}\} $, known as principal classes.
If all principal classes are excluded, the canonical dispersion will have the common factor $p_x^2$ or $p_y^2$, making $ p_x=0 $ or $p_y=0$ a critical line respectively.

Up to an exchange of $p_x$ and $p_y$, $ \{p_{x}^{m} p_{y}, p_{y}^n\} $ and $ \{p_{x}^{m}, p_{x}p_{y}^n\} $ are equivalent and there are three inequivalent principal classes, which determine the possible scaling exponents of an isolated critical point as ($m,n\geqslant 2$)
\begin{eqnarray}\label{eq_sp1}
\{p_{x}^m, p_{y}^n\}&:&\quad a=\frac{1}{m},b=\frac{1}{n},\\\label{eq_sp2}
\{p_{x}^{m} p_{y}, p_{y}^n\}&:&\quad a=\frac{n-1}{mn},b=\frac{1}{n},\\\label{eq_sp3}
\{ p_{x}^{m}p_y, p_{x}p_{y}^{n}\}&:&\quad a=\frac{n-1}{mn-1},b=\frac{m-1}{mn-1}.
\end{eqnarray}
Notice that principal classes are necessary but not sufficient conditions for isolated critical points. Even with terms in principal classes, when the coefficients are fine tuned (e.g. forming a complete square), the critical point can also be not isolated.

To summarize, the general procedure to work out the canonical dispersion of an isolated critical point has two steps: We first choose a pair of scaling exponents $ (a,b) $ in Eqs. (\ref{eq_sp1}-\ref{eq_sp3}),  
and then work out all allowed terms in the canonical dispersion according to all integer solutions $(m,n)$ to the equation $ ma+nb=1 $, as shown in Fig. \ref{fig_mn}.

As another application of scaling property, in the following we will discuss the DOS of canonical dispersions.

\subsection{Density of states}
The critical point $\bm p=\bm 0$ of canonical dispersion corresponds to a DOS singularity at $E=0$. According to its definition
\begin{eqnarray}
\rho(\varepsilon)=\int\frac{d^2\bm p}{(2\pi)^2}\delta[\varepsilon -E(\bm p)],
\end{eqnarray}
the DOS of canonical dispersion is also scale invariant $ \rho(\lambda{\varepsilon})=\lambda^{a+b-1}\rho({\varepsilon}) $, and hence follows the power law $ \rho(\varepsilon)\propto |\varepsilon|^{\nu} $ with exponent $ \nu =a+b-1 $.

From the results of possible scaling exponents in (\ref{eq_sp1}-\ref{eq_sp3}), DOS exponent $ \nu $ can only take specific rational values in the range $ -1<\nu\leqslant 0 $.
When $\nu=0$, the critical point is ordinary, and
the DOS singularity at $E=0$ can be discontinuity (extrema) or logarithmic divergence (saddle points). When $ \nu<0 $, $E=0$ is a power-law divergent singularity of DOS. 
The exponent $ \nu=-1 $ can only be approached when $m,n\to\infty$. A given value of $\nu$ may correspond to more than one pair of scaling exponents.

Since the scaling parameter has to be positive $ \lambda>0 $, $\rho(\varepsilon)$ and $\rho(-\varepsilon)$ cannot be related by scaling property (\ref{eq_QH}). Instead, they can be related by the analyticity of Green's function $ G(\varepsilon)=\int_{\bm p}[\varepsilon -E(\bm p)]^{-1} $, whose imaginary part determines DOS $ \rho(\varepsilon)=-{\rm Im}G(\varepsilon+i0^+)/\pi $. Combine this property with scaling property, we can write DOS of the canonical dispersion in the following compact form
\begin{eqnarray}\label{eq_div}
\rho(\varepsilon)={\rm Re}(C\varepsilon ^{\nu})=|C| |\varepsilon|^{\nu}\times
\begin{cases}
\cos\theta, & \varepsilon<0\\
\cos(\theta +\nu\pi), & \varepsilon>0
\end{cases},
\end{eqnarray}
where $C=|C|e^{i\theta}$ is a complex number with $\varepsilon>0$ as the branch cut of $\varepsilon^{\nu}$.
For local minima, $ \theta=\frac{1}{2}\pi $, while for local maxima $\theta=(\frac{1}{2}-\nu)\pi$.
For a saddle point the particle-hole asymmetry ratio $ \eta\equiv\rho(-|\varepsilon|)/\rho(|\varepsilon|)=\cos\theta/\cos(\theta +\nu\pi) $ is well-defined and nonzero.

The ordinary saddle point (\ref{eq_os}) has $ \nu =0 $ and $\theta =0$, hence the DOS peak is logarithmically divergent and particle-hole symmetric.

The high-order saddle point (\ref{eq_hs}) has $ \nu =-\frac{1}{4} $ and $\theta =0$, hence the DOS peak is power-law divergent with exponent $-\frac{1}{4}$ but particle-hole \textit{asymmetric} with asymmetry ratio $ \eta=\sec(\nu\pi)=\sqrt{2} $ \cite{magic}.

\subsection{Perturbation}
%Suppose a canonical dispersion contains a monomial term $ (m,n) $ and has scaling exponents $(a,b)$. Then $ ma+nb=1 $ and the scaling exponent of the monomial perturbation $(m',n')$ is $ \gamma =m'a+n'b\neq  1 $. When $ m'<m,n'<n $, the perturbation is relevant, and when $ m'>m,n'>n $, the perturbation is irrelevant. In other cases one needs to compute $ \gamma $ explicitly to determine the relevance of the perturbation.

Next we add perturbations to the canonical dispersion and find out the evolution of critical points.
With irrelevant perturbations only, the critical point at $\bm p=\bm 0$ is not affected, since irrelevant perturbations become negligible when arrpoaching $ \bm p=\bm 0 $. As a result, the total DOS is asymptotically power-law divergent (\ref{eq_div}) as $ \varepsilon\to 0 $, up to a non-singular background contribution \cite{magic}.

With relevant perturbations only, the total dispersion (\ref{eq_cata}) can have $m\geqslant 1$ critical points, whose properties depend on coefficients $\bm h$.
If we allow $\bm h$ to vary freely (i.e. without constraints from symmetry etc.), then the maximal number $ \mu\equiv{\rm max}\{m\} $ of critical points in (\ref{eq_cata}) is only determined by scaling property of canonical dispersion \cite{Gilmore,Arnold1,Arnold2,Arnold3}
\begin{eqnarray}\label{eq_mu}
\mu =(a^{-1}-1)(b^{-1}-1).
\end{eqnarray}
We define the positive integer $ \mu $ as the multiplicity (also known as Milnor number) of the critical point (\ref{eq_QH}). From all the possible scaling exponents in the three principal classes (\ref{eq_sp1}-\ref{eq_sp3}), we can also work out the corresponding multiplicity in each class.
It can be found that a given multiplicity can correspond to more than one canonical dispersion, which can belong to the same or different principal classes.

We find $ \mu =1 $ for the ordinary critical point (\ref{eq_os}), namely ordinary extrema and saddle points are robust against perturbations and does not need addtional tuning of the band structure \cite{VHS}.
For high-order critical point (\ref{eq_hs}), $ \mu =3 $, and under perturbations it will split into at most three ordinary critical points. When high-order critical point (\ref{eq_hs}) is at a generic point, one needs to tune 2 parameters to realize it, while at a mirror-invariant point only 1 parameter is needed, such as the twist angle in twisted bilayer graphene \cite{magic}.

For a critical point with multiplicity $\mu$, the number of independent relevant perturbations is at most $ \mu -1 $, since this critical point will split into at most $\mu$ critical points.
%The number and forms of independent relevant perturbations are determined by the canonical dispersion and symmetry group of the critical point. For example, a simple critical point at a generic point with trivial little group can have $ \mu -1 $ independent relevant perturbations. 
Details and examples of relevant perturbations will be discussed in subsection \ref{sp} and \ref{ss}.

At last we discuss the relation between topological index and multiplicity of an isolated critical point.
A critical point with topological index $I$ and multiplicity $\mu$ will eventually split into at most $e$ ordinary extrema and $s$ ordinary saddle points under perturbations,
which obey $ e+s=\mu $ and $ e-s=I $. To ensure that $e,s$ are both positive integers, $\mu$ and $I$ should have the same parity and $ \mu\geqslant |I| $. In fact this inequality can be stronger \cite{topology}
\begin{eqnarray}\label{eq_lm}
I\leqslant 1,\quad I^2\leqslant\mu,\quad  I=\mu ({\rm mod}2).
\end{eqnarray}
And reversely, if two integers $ \mu,I $ satisfy the {\it compatibility conditions} above, there always exists a critical point with topological index $I$ and multiplicity $\mu$.

We then apply Eq. (\ref{eq_lm}) to specific cases. When $\mu =1$, from conditions (\ref{eq_lm}) the topological index can only be $I=\pm 1$, namely only ordinary critical points in (\ref{eq_os}) are robust against perturbations.
When $\mu =3$, the topological index can only be $I=\pm 1$, namely only high-order critical points in (\ref{eq_hs}) can split into three ordinary critical points ($ e=1,s=2,I=-1 $ or $ e=2,s=1,I=+1 $).

The multiplicity of a local extremum is always odd since its topological index $ I=1 $ is odd, while the multiplicity of a saddle point can be any non-positive integer. These results will find explicit applications in the following subsections when we discuss specific examples of critical points.

%\subsection{Classification Schemes}
%which restrict coefficients in the Taylor expansion and can force some of them to be zero. In the following we consider mirror and in-plane rotation symmetries in 2D.
%
%When $ G=\{1,M\} $ with mirror symmetry $ M=M_{x}: p_{x}\to -p_{x} $ or $ M=M_{y}: p_{y}\to -p_{y} $, only even order terms in $p_x$ or $p_y$ are allowed respectively.
%
%When $ G=\{1,{\rm C}_2\} $ with twofold in-plane rotation symmetry $ {\rm C}_2:\bm p\to -\bm p $, the only allowed terms are $ p_{x}^np_{y}^{2m-n} $ with integers $ m,n $ and $ 2m\geqslant n $, namely all odd order terms are prohibited.
%
%When $ G=\{1,{\rm C}_n,\dots,{\rm C}_n^{n-1}\} $ with more than twofold in-plane rotation symmetry C$_n (n=3,4,6)$, the linear order terms vanish and the general dispersion reads
%\begin{eqnarray}\label{eq_cc}
%\mathcal{E}=\sum_{m=1}^{\infty}\left[\alpha_{m}p^{2m}+r_{m}p^{mn}\cos(mn\varphi +\varphi_m)\right]
%\end{eqnarray}
%where $ (p_x,p_y)=p(\cos\varphi,\sin\varphi) $. This expression will be mostly applied in Sec. \ref{ss} when we consider critical points at symmetric points.
%
%We can also include both $ {\rm C}_n $ rotation and mirror symmetry $M$ to form point group C$_{nv}$, which is equivalent to D$_n$ point group in 2D. Thus when considering symmetries in Secs. \ref{sp}, \ref{ss} and \ref{mod}, we will focus on point groups C$_{n}$ and C$_{nv}$, where C$_{1v}=\{1,M\}$ includes mirror symmetry $M$.

\subsection{Simple versus complicated critical points}\label{sp}
We can introduce two different classification schemes for critical points in 2D.
One way is to classify the critical point by DOS exponent. Notice that in 1D, the DOS of an ordinary critical point is already power-law divergent with exponent $-\frac{1}{2}$. When $ \nu>-\frac{1}{2} $, the critical point is called \textit{simple}, and when $\nu\leqslant -\frac{1}{2}$ the critical point is complicated.

We find three classes of simple critical points in 2D: $A$ class with $b=\frac{1}{2}$, $D$ class with $ 2a+b=1 $, and $E$ class with $ b=\frac{1}{3} $, up to an exchange of $a$ and $b$. Among them, $A$ and $D$ classes are infinite series, while $E$ class consists of three inequivalent critical points. In the following we will derive the canonical dispersions of simple critical points, depending on the topology and scaling.

%\subsection{Generic position}
%We first consider the {simple} critical points at {generic} positions with no extra symmetry, whose canonical dispersions can be constructed by all integer solutions to $ma+nb=1$.
%As shown in Sec. \ref{os} and the Appendix, we can always find an invertible and smooth coordinate transform (which can be nonlinear) in both momentum and energy, under which the canonical dispersion includes only two terms and becomes parameter-free.
%In the following, we write down such canonical dispersions of simple critical points, according to the multiplicity $\mu$, topology and $A,D,E$ classes given above.

The saddle point in $ A $ class can have any integer multiplicity $ \mu\geqslant 1 $ with canonical dispersion
\begin{eqnarray}\label{eq_A}
A_{\mu}=p_{x}^{\mu +1}- p_{y}^2,
\end{eqnarray}
while the local extremum in $A$ class can only have odd multiplicity $ \mu=2n-1 $ ($ n\geqslant 1 $) with canonical dispersion
\begin{eqnarray}\label{eq_Ap}
A'_{\mu}=p_{x}^{\mu +1}+ p_{y}^2.
\end{eqnarray}
The odd multiplicity of a local extrmum is due to constraint (\ref{eq_lm}) as discussed in the last subsection.

\begin{figure*}
\centering
\subfloat{{\includegraphics[width=18cm]{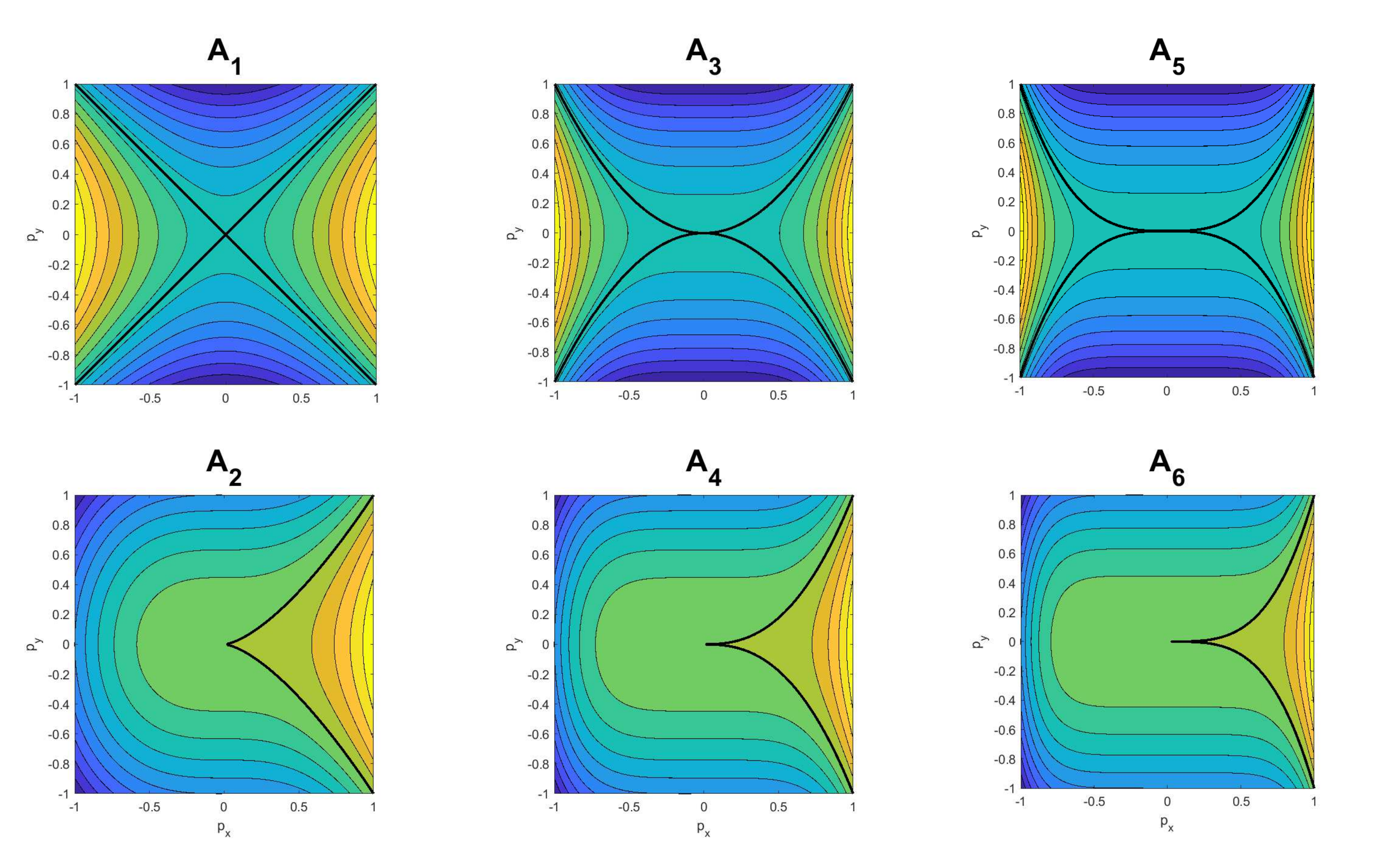} }}
\qquad
\subfloat{{\includegraphics[width=18cm]{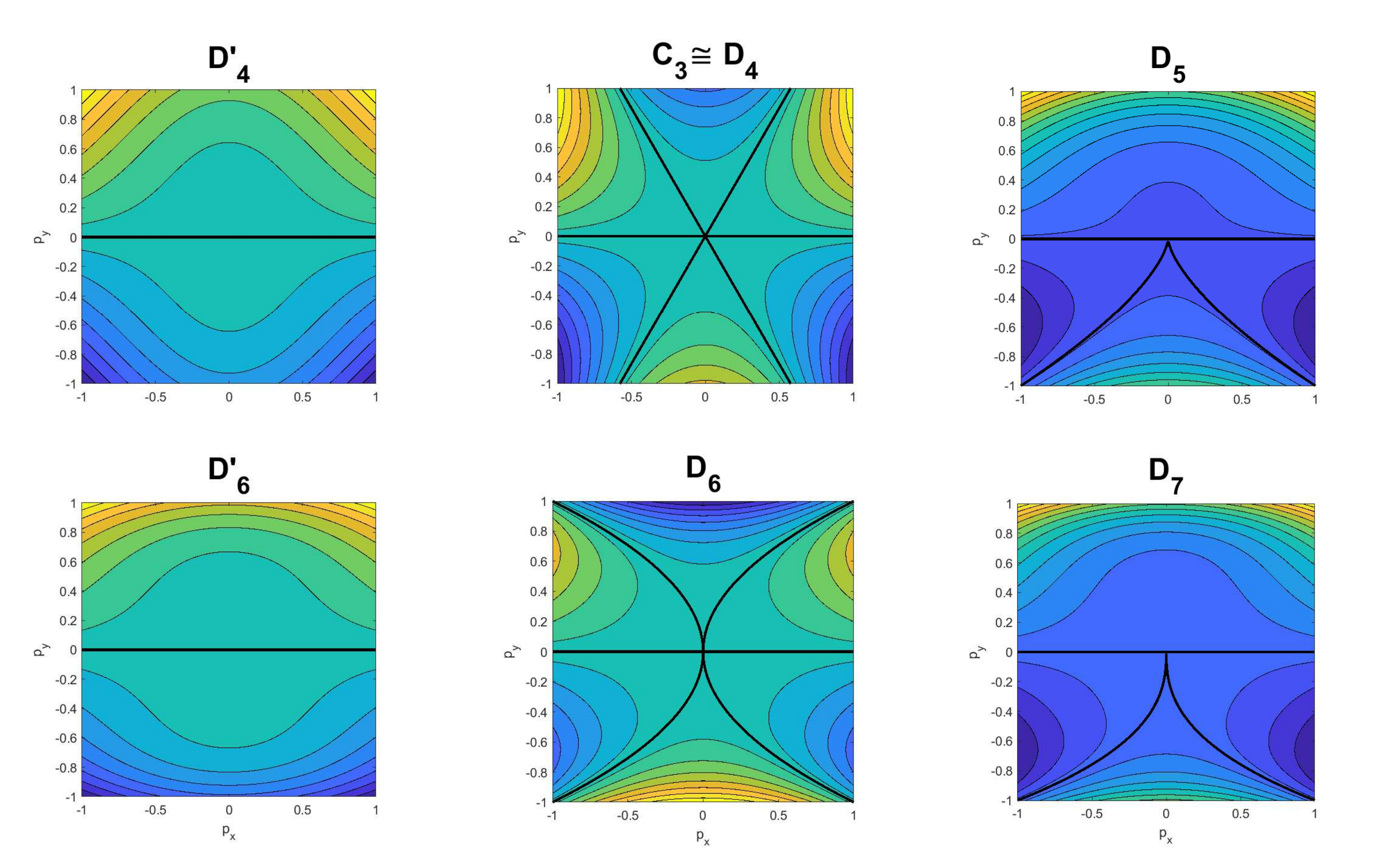} }}
\caption{Energy contours of $A$ and $D$ classes, where the origin is the critical momentum point, colors denote energy (orange is higher and blue is lower) and black lines are energy contours at critical point energy. Here $C_3={\rm Re}[(p_x+ip_y)^3]$, $D_4=p_{x}^2p_{y}-p_{y}^3$ and $ C_3\cong D_4 $ means they are equivalent up to a linear coordinate transform.}
\label{fig_ad}
\end{figure*}

Notice that $A_1,A'_1$ describe ordinary critical points (\ref{eq_os}) and (\ref{eq_osc}), and $A_3,A'_3$ are the high-order critical points in (\ref{eq_hs}) and (\ref{eq_hsc}).
As shown in Fig. \ref{fig_ad}, $A_{2n}$ denotes a beak point of a single energy contour, with topological index 0. While $A_{2n-1}$ denotes a merging point of two energy contours, with topological index $-1$.

The simple saddle point in $D$ class with multiplicity $\mu$ has one of the following two canonical dispersions
\begin{eqnarray}\label{eq_D}
D_{\mu}=p_{x}^2p_{y}-p_{y}^{\mu -1},\quad D'_{\mu}=p_{x}^2p_{y}+p_{y}^{\mu -1}.
\end{eqnarray}
Notice that $ D_{2n-1} $ and $ D'_{2n-1} $ are equivalent upon coordinate transform $ p_{y}\to -p_{y} $.

As shown in Fig. \ref{fig_ad}, $D_{2n}$ describes three energy contours merging at the singular point with topological index $-2$.
On the other hand, $D_{2n-1}$ describes one straight energy contour meeting with another energy contour at its beak point, and the topological index is $-1$.
Notice that for low multiplicity $\mu\leqslant 3$, $ D_1 $ describes a singular line $p_x=0$ instead of an isolated singular point, $D_2$ is regular at $\bm p=\bm 0$, and $D_3$ is equivalent to $ A_3 $ according to scaling property as shown in the Appendix \ref{ct}. In the following we always require $ \mu\geqslant 4 $ in $D$ class.

\begin{figure*}
\includegraphics[width=1\textwidth]{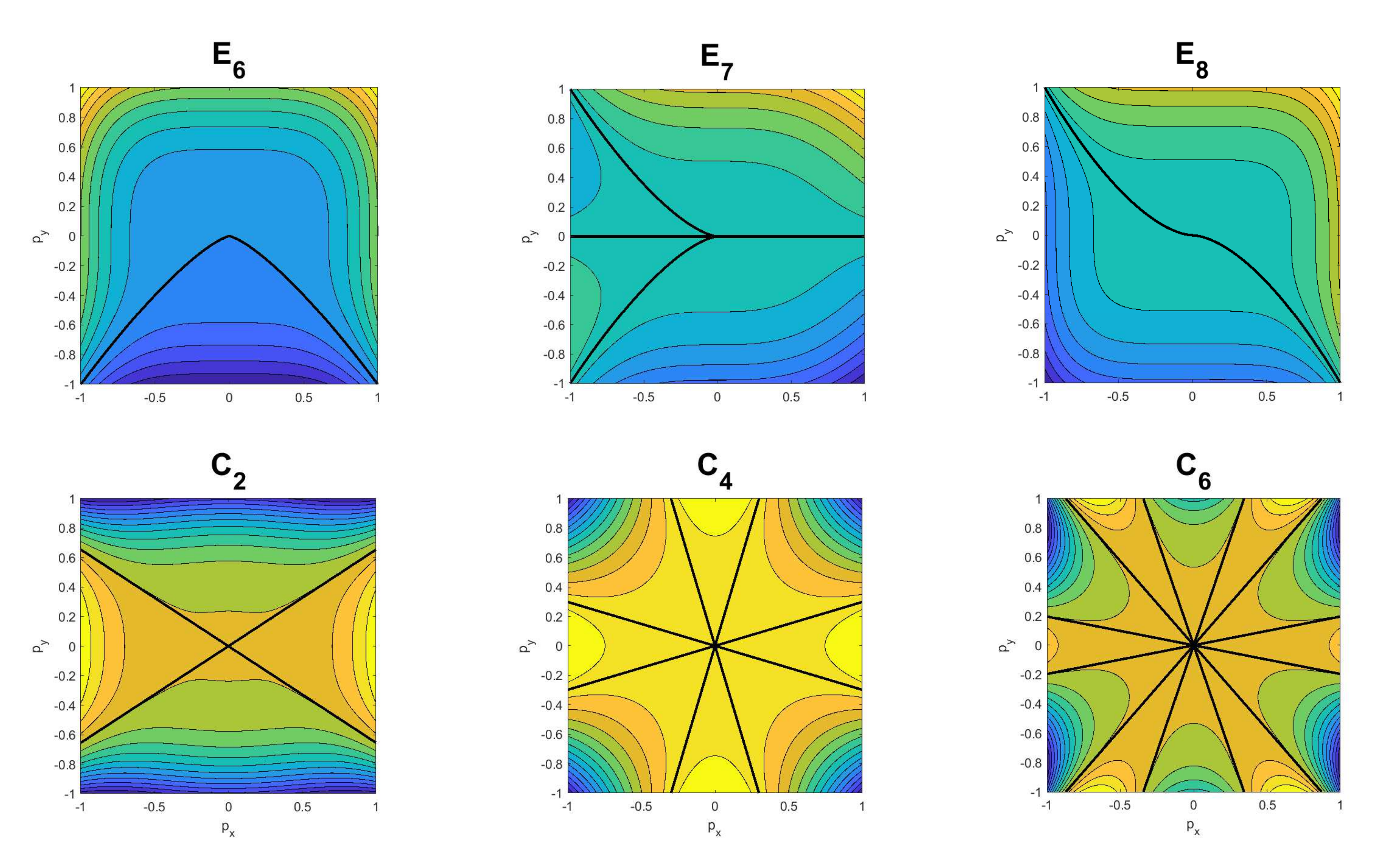}
\centering
\caption{{Energy contours of $E$ and $C$ classes, where the origin is the critical momentum point and black lines are energy contours at critical point energy. The continuous parameters of all three $C$ classes are the same $ r=-0.4 $.}}\label{fig_ec}
\end{figure*}

Critical points in $E$ class are all saddle points, whose the multiplicity can only be $ \mu =6,7,8 $ with canonical dispersions
\begin{equation}\label{eq_E}
E_{6}=p_{x}^4+p_{y}^3,\quad
E_{7}=p_{x}^3p_{y}+p_{y}^3,\quad
E_{8}=p_{x}^5+p_{y}^3,
\end{equation}
and the topological index is $ I(E_\mu)=\frac{1}{2}[(-)^\mu-1] $. 
%All $E$ class critical points are saddle points with 
Energy contours of $E$ class are shown in Fig. \ref{fig_ec}.

For simple saddle points, the topological index, multiplicity, DOS exponents and asymmetry property of saddle points are summarized in Table. \ref{S}. We find that scaling exponents and topological index can uniquely determine the dispersion of a simple critical point up to a coordinate transform, just like 1D cases in Sec. \ref{1d}. In fact, 1D cases are equivalent to $A$ class.

\begin{figure}
\includegraphics[width=0.45\textwidth]{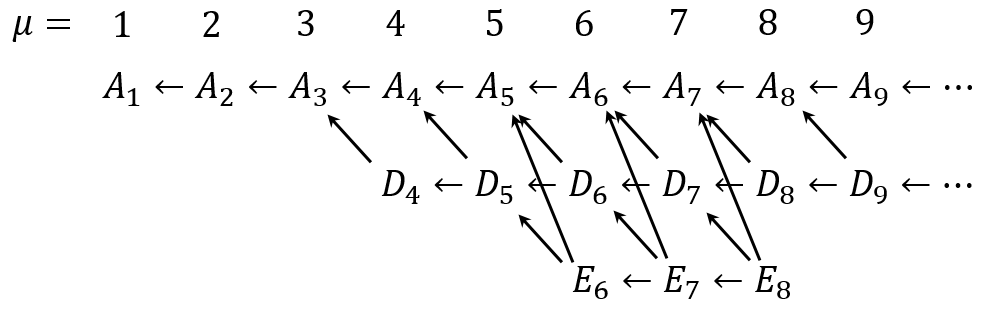}
\centering
\caption{{Hierarchy diagram of simple critical points (including saddle points and local extrema). Under relevant perturbation, a simple critical point can split to simple critical points with lower multiplicities indicated by the arrows. Here $A,D$ classes also include variants $A',D'$.}}\label{fig_2}
\end{figure}

Under relevant perturbations listed in Table. \ref{T}, a simple critical point can split into simple critical points with lower multiplicities through a topological transition \cite{Gilmore,Arnold1,Arnold2,Arnold3}. The intermediate and final products of such topological transitions are summarized in Fig. \ref{fig_2}.

Unlike simple critical points, complicated critical points cannot be fully classified. In the next subsection, we will list some examples of complicated critical points at symmetric positions.

\begin{table*}
\setlength{\tabcolsep}{12pt}
\renewcommand{\arraystretch}{2}
\centering
\begin{tabular}{c|cccc} \hline
Critical point & Topological index & Multiplicity & DOS exponent & DOS asymmetry\\%[0.2cm]
\hline\hline
%$A_{1} $ & $ p_{x}^{2}-p_{y}^2 $   & $ - 1 $ & $ 1 $ & $ 0 $(log)\\[0.2cm]\hline
$A_{n} $ ($n\geqslant 1$) & $ \displaystyle -\frac{1}{2}[1-(-)^n] $ & $ n $ & $ \displaystyle\frac{1}{n+1}-\frac{1}{2} $ &
$\displaystyle \theta=\frac{-\pi}{2n+2} $ (even), $ \displaystyle \theta=0 $ (odd) \\[0.2cm]
\hline
$D_{n} $ ($n\geqslant 4$) & $ \displaystyle -\frac{1}{2}[3+(-)^n] $ & $ n $ & $ \displaystyle\frac{1}{2(n-1)}-\frac{1}{2} $ & $ \eta=1 $ (even), $ \displaystyle \theta=\frac{\pi}{4} $ (odd) \\[0.2cm]
\hline
$D'_{2n} $ ($n\geqslant 2$) & $ \displaystyle 0 $ & $ 2n $ & $ \displaystyle\frac{1}{2(2n-1)}-\frac{1}{2} $ & $ \eta =1 $ \\[0.2cm]
\hline
$E_{6} $ & $ 0 $ & $ 6 $ & $ \displaystyle-\frac{5}{12} $ & $ \displaystyle \theta=\frac{\pi}{12} $ \\[0.2cm]
\hline
$E_{7} $ & $ -1 $ & $ 7 $ & $ \displaystyle-\frac{4}{9} $ & $ \eta=1 $ \\[0.2cm]
\hline
$E_{8} $ & $ 0 $ & $ 8 $ & $ \displaystyle-\frac{7}{15} $ & $ \eta=1 $ \\[0.2cm]
\hline\hline
$C_{2} $ & $ -1 $ & $ 9 $ & $\displaystyle-\frac{1}{2}$ & $ \displaystyle\theta =-{\rm Arg}{\rm K}\left(s\right) $\\[0.2cm]
\hline
$C_{n} $ ($ n=4,6 $) & $ 1-n $ & $ (n-1)^2 $ & $\displaystyle \frac{2}{n}-1$ & $ \displaystyle\theta =\left(\frac{1}{2}-\frac{2}{n}\right)\pi-{\rm Arg}{F}_{n}\left(s\right) $ \\[0.2cm]
\hline
\end{tabular}
\caption{Properties of saddle points in classes $ A,D,D',E,C $. In classes $A$ and $D$, even/odd denotes the pairity of integer $n$, the asymmetry of the DOS peak is either described by angle $\theta$ or $ \eta=\cos\theta/\cos(\theta +\nu\pi) $. In classes $C$, $ |r|<1 $, Arg$ z $ is the phase of complex number $z$, $ {\rm K}(s) $ is the complete elliptic integral of the first kind, $ F_{n}(s)\equiv _{2}{F}_{1}\left(\frac{2}{n},\frac{1}{2},1; s\right) $ is the hypergeometric function and the argument is $ s=2/(r+1) $. the DOS exponent also applies to local extrema of the same class.}
\label{S}
\end{table*}

\begin{table*}
\setlength{\tabcolsep}{12pt}
\renewcommand{\arraystretch}{2}
\centering
\begin{tabular}{c|ccc} \hline
Critical point & Canonical dispersion & Relevant perturbation & Scaling exponents\\%[0.2cm]
\hline\hline
$A_{n} $ ($n\geqslant 1$) & $ p_{x}^{n+1}\pm p_{y}^2 $ & $ \displaystyle\bm H=\bm H(M_{y})=\{p_{x}^j|1\leqslant j\leqslant n-1\}  $ & $ \displaystyle a=\frac{1}{n+1},\quad b=\frac{1}{2}  $\\[0.2cm]
 &  & $ \displaystyle\bm H(M_{x})=\{p_{x}^{2j}|1\leqslant j\leqslant\frac{1}{2}(n-1)\}  $ ($n$ is odd) & 
\\[0.2cm]
\hline
$D_{n} $ ($n\geqslant 4$) & $ p_{x}^2p_{y}\pm p_{y}^{n-1} $  & $ \displaystyle\bm H=\{p_{x},p_{x}^2,p_{y}^j|1\leqslant j\leqslant n-3\} $ & $ \displaystyle a=\frac{n-2}{2(n-1)},\quad b=\frac{1}{n-1}  $\\[0.2cm]
 &  & $ \displaystyle\bm H(M_{x})=\{p_{x}^2,p_{y}^j|1\leqslant j\leqslant n-3\} $ \\[0.2cm]
$D_{4}\cong C_3 $ & $ 3p_{x}^2p_{y}- p_{y}^{3} $  & $ \bm H({\rm C}_{3})=\bm H({\rm C}_{3v})=\{p^2\} $ & $ \displaystyle a=b=\frac{1}{3}  $\\[0.2cm]
\hline
$E_{6} $ & $ p_{x}^4+p_{y}^3 $ & $ \displaystyle\bm H=\{p_{x},p_{x}^2,p_{y},p_{x}p_{y},p_{x}^2p_{y}\} $ & $ \displaystyle a=\frac{1}{4},\quad b=\frac{1}{3}  $\\[0.2cm]
  & & $ \displaystyle\bm H(M_{x})=\{p_{x}^2,p_{y},p_{x}^2p_{y}\} $ \\[0.2cm]
\hline
$E_{7} $ & $ p_{x}^3p_{y}+p_{y}^3 $ & $ \displaystyle\bm H=\{p_{x},p_{x}^2,p_{x}^3,p_{x}^4,p_{y},p_{x}p_{y}\} $ & $ \displaystyle a=\frac{2}{9},\quad b=\frac{1}{3}  $\\[0.2cm]
\hline
$E_{8} $ & $ p_{x}^5+p_{y}^3 $ & $ \displaystyle
\bm H=\{p_{x},p_{x}^2,p_{x}^3,p_{y},p_{x}p_{y},p_{x}^2p_{y},p_{x}^3p_{y}\} $ & $ \displaystyle a=\frac{1}{5},\quad b=\frac{1}{3}  $\\[0.2cm]
\hline\hline
$C_{2} $ & $ {\rm Re}[(p_{x}+ip_{y})^{2}]p^2+rp^4 $ & $\displaystyle\bm H({\rm C}_2)=\{p_{x}^2,p_{y}^2,p_{x}p_{y}\}$ & $ \displaystyle a=b=\frac{1}{4}  $\\[0.2cm]
 &  & $\displaystyle\bm H({\rm C}_{2v})=\{p_{x}^2,p_{y}^2\}$\\[0.2cm]
\hline
$C_{n} $ ($ n=4,6 $) & $ {\rm Re}[(p_{x}+ip_{y})^{n}]+rp^{n} $ & $\displaystyle \bm H({\rm C}_n)=\bm H({\rm C}_{nv})=\{p^{2j}|1\leqslant j\leqslant \frac{n}{2}-1\}$ & $ \displaystyle a=b=\frac{1}{n}  $\\[0.2cm]
\hline
%$O$ & $p^4$ & $ \displaystyle \bm H({\rm C}_3)=\bm H({\rm C}_{3v})=\{p^2,3p_{x}^2p_{y}-p_{y}^3\} $& $ \displaystyle a=b=\frac{1}{4}  $\\[0.2cm]
%  &   & $ \displaystyle \bm H({\rm C}_6)=\bm H({\rm C}_{6v})=\bm H({\rm O}(2))=\{p^2\} $\\[0.2cm]
\hline
\end{tabular}
\caption{Canonical dispersions, relevant perturbations and scaling exponents of critical points in classes $ A,D,E,C$, where $C$ has a single continuous parameter $r\neq\pm 1$. In Perturbation, $ \bm H(G) $ denotes the linearly independent perturbations allowed at the momentum point with symmetry group $G$. When $ G=\{1\} $ we denote $\bm H(G)$ simply as $\bm H$. As mentioned previously, $\bm H$(D$_{n})=\bm H($C$_{nv}$) in 2D.}
\label{T}
\end{table*}

\subsection{Symmetric versus generic critical points}\label{ss}
Another way is to classify the critical point by symmetry. At a momentum point with little group $G$, the dispersion has to satisfy symmetry constraints
\begin{eqnarray}
\mathcal{E}({g\bm p})=\mathcal{E}({\bm p})\quad (\forall g\in G).
\end{eqnarray} 

%The two classification schemes (by DOS exponent and by symmetry) are not parallel, and the same critical point may be labeled differently in different schemes. In such cases, we use $ \cong $ to denote the equivalent critical points in two classification schemes, for example $ D_4\cong C_3 $ as shown later.

%Next we consider critical points at symmetric positions. As mentioned previously, we will focus on mirror and in-plane rotation symmetries in the following.

%When a critical point is invariant under mirror symmetry $ M_{y}: p_{y}\to -p_{y} $, the simple critical point can still belong to $A_{\mu}(A'_{\mu})$ class, and the number of allowed perturbations is still $ \mu -1 $.

When a critical point is invariant under mirror symmetry $ M: p_{i}\to -p_{i} $ ($i=x$ or $y$), the simple critical point belongs to either $A,A'$ class, $D,D' $ classes or $E_6$ class.
We find that some simple critical points can still satisfy mirror symmetry, but the allowed perturbations may be restricted as listed in Table. \ref{T}.

When the critical point has twofold in-plane rotation symmetry $ {\rm C}_2:\bm p\to -\bm p $, simple critical points can only belong to $A_{2n-1}$ or $A'_{2n-1}$ class ($n\geqslant 1$). In the presence of more than twofold in-plane rotation symmetry, when second order terms do not vanish, we find the critical point can only be an ordinary energy extremum $A'_1$, independent of in-plane rotation symmetry.

Without second order terms and with $ {\rm C}_3 $ symmetry, there can be simple saddle point $ C_{3}={\rm Re}[(p_{x}+ip_{y})^{3}] $, which is equivalent to $ D_4 $ up to a linear coordinate transform, denoted as $C_3\cong D_4$. When second order terms vanish while $ {\rm C}_4 $ or $ {\rm C}_6 $ symmetry is present, no simple critical point can exist. 
As shown in Appendix \ref{mod}, one can realize $ C_3 $ saddle point at $\Gamma$ or $K$ points in the BZ of a triangular lattice.

The symmetry-restricted relevant perturbations of simple critical points are summarized in Table. \ref{T}.

%\section{Rotation-symmetric critical points}\label{ss}
Besides simple critical points, we can consider the complicated critical point whose dispersion is a homogeneous polynomial of order $N$ and symmetric under in-plane rotations. The scaling exponents in this case are $ a=b=1/N $ and the multiplicity is $ \mu =(N-1)^2 $. When $ N $ is even, the critical point can be a saddle point or a local extremum, while for odd $N$ the critical point can only be a saddle point. In the special case of $N=2$, the homogeneous critical point is ordinary, and when $N=3$ a simple saddle point $C_3\cong D_4$. In the following, we consider complicated critical points with $ N\geqslant 4 $ which can be usually realized at rotation-invariant momentum points.

When the critical point has $ {\rm C}_2 $ symmetry, the homogeneous dispersion can have order $N=4$
\begin{eqnarray}\label{eq_C2}
C_{2}={\rm Re}[(p_{x}+ip_{y})^{2}]p^{2}+rp^{4},
\end{eqnarray}
and contains a continuous parameter $ |r|\neq 1 $, where $p^2=p_x^2+p_y^2$. The multiplicity is 9, and the topological index is $ I={\rm sgn}(|r|-1) $. When $ r=+1 $ or $r=-1$ there will be a critical line $ p_x=0 $ or $ p_y=0 $ respectively.

With $ {\rm C}_{n} $ symmetry ($ n=4,6 $), the homogeneous dispersion can have order $N=n$
\begin{eqnarray}\label{eq_Cn}
C_{n}={\rm Re}[(p_{x}+ip_{y})^{n}]+rp^{n},
\end{eqnarray}
with a continuous parameter $ |r|\neq 1 $ and odd multiplicity $ \mu =(n-1)^2 $. When $ |r|>1 $, the topological index is $I=1$. When $|r|<1$ the topological index is $I=1-n<0$ and $ \mu =I^2 $ holds. When $ |r|=1 $ there will be $\frac{1}{2}n$ critical lines passing through the critical point with polar angles $\varphi$ satisfying $ \cos n\varphi =-r $.

As shown in the second line of Fig. \ref{fig_ec}, when $|r|<1$, the saddle point $C_{n} (n=2,4,6)$ describes an intersecting point of $ n $ energy contours.

Critical points in the finite series of $C_{n}$ $(n=2,3,4,6) $ class can also exist at momentum points with symmetry group $G\subset {\rm C}_{n}$ or ${\rm C}_{nv}$, while the allowed perturbations may be restricted. The canonical dispersion and allowed perturbations for different symmetry groups are listed in Table. \ref{T} for $C$ class critical points. The topological index, multiplicity, DOS exponents and asymmetry are summarized in Table. \ref{S} for $C$ class saddle points.

%With $ {\rm C}_3 $ symmetry, the critical point can also be a complicated local extremum $O=p^4,$ and only two independent relevant perturbations $p^2$ and $3p_{x}^2p_{y}-p_{y}^3$ are allowed by symmetry. In the special case with only $p^2$ perturbation, the energy dispersion $ E=p^4+hp^2 $ has a critical ring with radius $ p=\sqrt{-h/2} $ when perturbation $h<0$, and the critical points are not isolated. Hence the multiplicity of critical point $O$ is formally divergent.
%Critical point $O$ is the special limit of $C_2$ and $C_4$ when $ |r|\to\infty $, and has full rotation symmetry O(2). Thus it can also exist at momentum points with any symmetry group $ G\subset $O(2). The canonical dispersion and allowed perturbations for different symmetry groups are also listed in Table. \ref{T} for critical point $O$.

\begin{figure}
\includegraphics[width=0.45\textwidth]{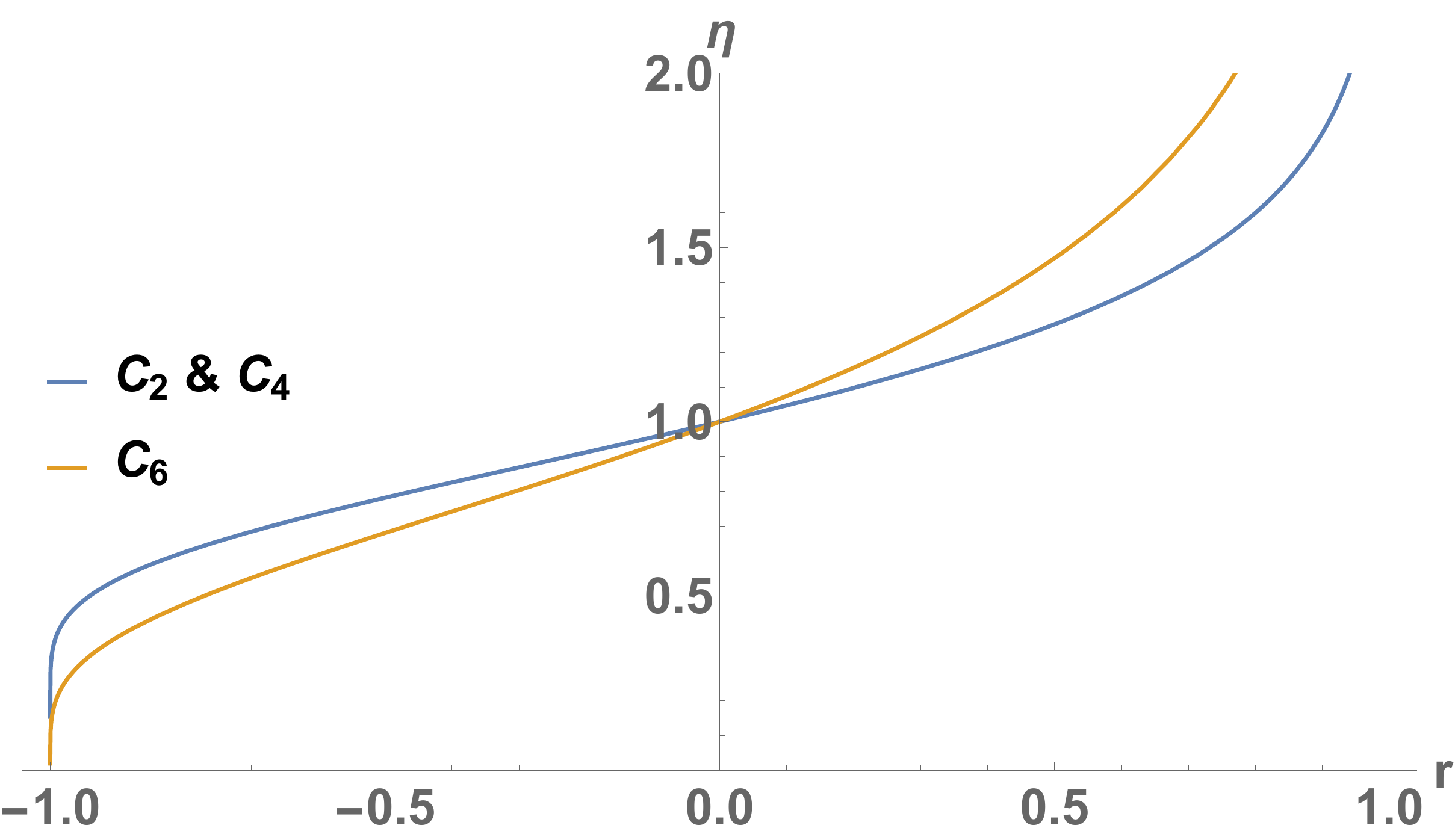}
\centering
\caption{{DOS asymmetry ratio $\eta$ of $C_2,C_4$ and $C_6$ as functions of the continuous parameter $r\in (-1, 1)$.}}\label{fig_eta}
\end{figure}

We find that critical points $C_2,C_4$ have the same scaling exponents $a=b=\frac{1}{4}$. Local extrema $C_2,C_4$ have the same topological index $I=1$, saddle points $C_2,C_4$ have different topological indices but the same DOS asymmetry parameter as shown in Fig. \ref{fig_eta} together with that of $C_6$.

In the following, we introduce the general method to describe critical points in 3D, also based on general principles of topology, scaling and symmetry.

\section{High-Order Critical Points in 3D}\label{g3d}

\subsection{Topology}
To describe the topology of a critical point in 3D, we can introduce the {topological index}
\begin{eqnarray}\label{eq_pi3D}
I\equiv\frac{1}{4\pi}\oiint_{\mathcal{S}}\hat{\bm v}\cdot\frac{\partial\hat{\bm v}}{\partial p_{\parallel}}\times\frac{\partial\hat{\bm v}}{\partial p_{\perp}}dp_{\parallel}dp_{\perp}
\end{eqnarray}
as the winding number of the group velocity $ \bm v=\nabla_{\bm p}\mathcal{E} $ along a closed surface $\mathcal{S}$ which encloses $ \bm p=\bm 0 $ as the only critical point, where $ p_{\perp,\parallel} $ are orthogonal coordinates on the surface and $\hat{\bm v}=\bm v/|\bm v|$. In order to make topological index well-defined, $ \bm p=\bm 0 $ has to be the only critical point in its neighborhood, which is known as an isolated critical point. When we choose spherical coordinates $ p_{\perp}=\theta,p_{\parallel}=\phi $, the winding direction of $\bm v$ is chosen as $ \theta:-\pi\to\pi $ and $ \phi:0\to 2\pi $. 

As a concrete example, we consider an ordinary critical point in 3D whose dispersion reads $E=\bm p^{\rm T}D\bm p$ with Hessian matrix $D$. We work out the topological index $ I={\rm sgn}(\det D) $. Hence a local minimum has topological index $ I=+1 $ and a local maximum has $ I=-1 $. For a saddle point, the topological index can either be $ I=+1 $ or $ I=-1 $.

In general the topological index of an isolated critical point is
\begin{eqnarray}\label{eq_top3D}
I=n_{e}-n_{h}
\end{eqnarray}
where $ n_e (n_h) $ is the number of patches of the energy surface at $E>0$ ($E<0$) within the region enclosed by $\mathcal{S}$. 
From its definition (\ref{eq_pi3D}), we find the topological index in 3D is odd under energy inversion $ E\to -E $, which is reflected by the odd parity of Eq. (\ref{eq_top3D}) under exchange $ n_e\leftrightarrow n_h$.

To prove this formula, we notice that the solid angle of velocity vector winding around an energy surface is $ \pm(4\pi -\Omega) $, where $ \Omega $ is the total solid angle of open regions and $ \pm $ applies to electron (hole) surface. Thus the total solid angle of velocity vector winding around the surface $ \mathcal{S} $ is $ \Omega =4\pi(n_e-n_h)-(\Omega_e-\Omega_h) $ where $\Omega_{e(h)}$ is the total solid angle of open regions in electron (hole) surfaces. Notice that by continuity, the electron and hole surfaces will share the same open regions $ \Omega_e=\Omega_h $, and the topological index is found to be $ I=\Omega/(4\pi)=n_e-n_h $.

In 3D, we have the similar Poincar\'e-Hopf theorem
\begin{eqnarray}\label{eq_charges}
\sum_{i} I_{i}=0.
\end{eqnarray}
Due to the finite and periodic BZ, the smooth and bounded energy dispersion will have at least one minimum and one maximum. According to Eq. (\ref{eq_charges}) there can be no saddle points in 1D and 3D. Unlike 1D, the dispersion in 3D without saddle points usually requires additional tuning.

\subsection{Scaling}
The discussion of scaling properties in 2D can be generalized to 3D and also 1D without much modifications. The way we define canonical dispersion and perturbations will be the same, while the exponents are slightly modified.

The {canonical} dispersion in 3D also has vanishing linear terms $ \nabla E|_{\bm p=\bm 0}=\bm 0 $ and is scale invariant
\begin{eqnarray}\label{eq_QHs}
E(\lambda^{a} p_{x},\lambda^{b}p_{y},\lambda^{c}p_{z})=\lambda E(p_{x},p_{y},p_z),
\end{eqnarray}
with three scaling exponents $ a,b,c>0 $, where $\lambda>0$ is an arbitrary positive number. 
The multiplicity and DOS exponent of the canonical critical point are
\begin{eqnarray}
\mu &=&(a^{-1}-1)(b^{-1}-1)(c^{-1}-1),\\
\nu &=&a+b+c-1.
\end{eqnarray}

The canonical dispersion of an isolated critical point has to contain at least one of the seven principal classes listed in Table. \ref{S1} to avoid critical lines and critical planes, and hence all the possible scaling exponents are exhaustively worked out in Table. \ref{S1}.
For the same reason as 2D case, principal classes are necessary but not sufficient for isolated critical point.

In 3D, due to similar reasons, we can derive the {\it compatibility conditions}
\begin{eqnarray}\label{eq_cp}
|I|\leqslant\mu,\quad  I=\mu ({\rm mod}2).
\end{eqnarray}

\begin{table*}
\setlength{\tabcolsep}{12pt}
\renewcommand{\arraystretch}{2}
\centering
\begin{tabular}{c|c} \hline
Principal Class & Scaling Exponents %& Multiplicity $ \mu $
\\%[0.2cm]
\hline\hline
$\{p_{x}^{m},p_{y}^{n},p_{z}^{l}\}$&$\displaystyle a=\frac{1}{m},b=\frac{1}{n},c=\frac{1}{l} $
%&$\displaystyle (m-1)(n-1)(l-1)$
\\[0.2cm]
\hline
$\{p_{x}^{m},p_{y}^{n},p_{z}^{l}p_y\}$&$\displaystyle a=\frac{1}{m},b=\frac{1}{n},c=\frac{n-1}{nl} $%&$\displaystyle (m-1)(nl-n+1)$
\\[0.2cm]
\hline
$\{p_{x}^{m},p_{y}^{n}p_x,p_{z}^{l}p_x\}$&$\displaystyle a=\frac{1}{m},b=\frac{m-1}{mn},c=\frac{m-1}{ml}$%&$\displaystyle \frac{(mn-m+1)(ml-m+1)}{m-1}$
\\[0.2cm]
\hline
$\{p_{x}^{m},p_{y}^{n}p_z,p_{z}^{l}p_y\}$&$\displaystyle a=\frac{1}{m},b=\frac{l-1}{nl-1},c=\frac{n-1}{nl-1}$\\[0.2cm]
\hline
$\{p_{x}^{m},p_{y}^{n}p_z,p_{z}^{l}p_x\}$&$\displaystyle a=\frac{1}{m},b=\frac{ml-m+1}{mnl-1},c=\frac{m-1}{ml-1}$\\[0.2cm]
\hline
$\{p_{x}^{m}p_y,p_{y}^{n}p_x,p_{z}^{l}p_x\}$&$\displaystyle a=\frac{n-1}{mn-1},b=\frac{m-1}{mn-1},c=\frac{(m-1)n}{(mn-1)l}$\\[0.2cm]
\hline
$\{p_{x}^{m}p_y,p_{y}^{n}p_z,p_{z}^{l}p_x\}$&$\displaystyle a=\frac{nl-l+1}{mnl+1},b=\frac{ml-m+1}{mnl+1},c=\frac{mn-n+1}{mnl+1}$\\[0.2cm]
\hline
\end{tabular}
\caption{Principal classes and corresponding scaling exponents for isolated critical points.}
\label{S1}
\end{table*}

From scaling exponents in Table \ref{S1}, we find DOS is not necessarily divergent as $ -1<\nu\leqslant\frac{1}{2} $. The 3D high-order critical point with divergent DOS $ (\nu\leqslant 0) $ should have even higher order than its 2D counterpart.
%\subsection{Trivial extension}
In fact, for a given critical point with canonical dispersion $E_{\rm 2D}(p_x,p_y)$, its {\it trivial extension} to 3D is given by
\begin{eqnarray}
E_{\rm 3D}(p_x,p_y,p_{z})=E_{\rm 2D}(p_x,p_y)+Wp_{z}^2,
\end{eqnarray}
with the following topological index, multiplicity and DOS exponent
\begin{eqnarray}
I_{\rm 3D}=I_{\rm 2D}\cdot{\rm sgn}W,\quad
\mu_{\rm 3D}=\mu_{\rm 2D},\quad
\nu_{\rm 3D}=\nu_{\rm 2D}+\frac{1}{2}.
\end{eqnarray}
By trivial extension, simple critical points will not produce divergent DOS but only complicated ones do.

In the following Sections we study critical points which can be realized by tuning one parameter, where we study 2D examples in Sec. \ref{2d}, and 3D in Sec. \ref{3d}.

\section{High-Order Critical points with single tuning parameter}\label{per}
In this Section we consider the high-order critical points which can be realized by tuning one parameter, which can be twist angle, strain, pressure or external fields. In 2D, they are $ A_2 $ at generic points, $ A_3,A_{3}^{\prime} $ at mirror-invariant points, and $C_3,C_4$ at rotation-invariant points. In 3D, we consider an example with cubic symmetry. In all these cases, the total topological index is always conserved.

\subsection{2D examples}\label{2d}
At generic points, only $A_2$ saddle point can be stably realized by tuning one parameter $h$
\begin{eqnarray}\label{eq_a2}
E =hp_{x}+p_{x}^3-p_{y}^2.
\end{eqnarray}
Under perturbation $h$, there will be at most 2 critical points according to the multiplicity of $ A_2 $. Since the topological index of $ A_2 $ is 0, there are either two ordinary critical points with opposite topological indices, or one $A_2$ critical point, or no critical points in Eq. (\ref{eq_a2}).

To see the critical points explicitly, we can compute DOS of dispersion (\ref{eq_a2}). The DOS of dispersion (\ref{eq_a2}) can be expressed in terms of the elliptic integral of the first kind K$(z)$ as
\begin{equation}\label{eq_dosa2}
\rho(\varepsilon)=\frac{1}{\pi^2}{\rm Re}\left\{\frac{1}{\sqrt{z_{1}-z_{2}}}{\rm K}\left(\frac{z_{1}-z_{0}}{z_{1}-z_{2}}\right)\right\}
\end{equation}
where $ z_{j}=\frac{1}{3}\omega^{j}f-h\omega^{-j}f^{-1} $ with $ \omega=e^{2\pi i/3} $ and $f=\left(\frac{3}{2}\sqrt{81\varepsilon^2+12h^3}-\frac{27}{2}\varepsilon\right)^{1/3}$.

\begin{figure}
\includegraphics[width=0.5\textwidth]{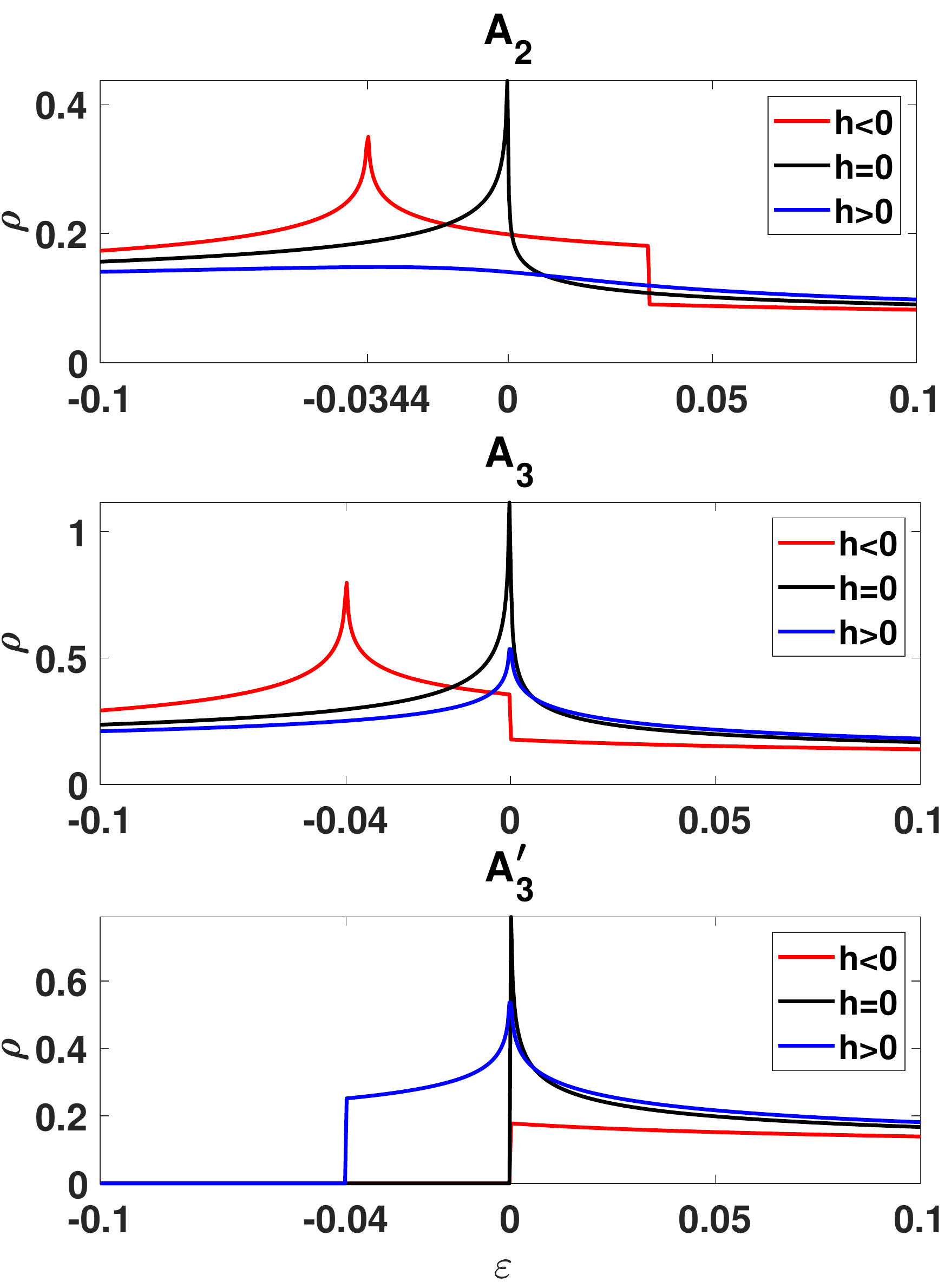}
\centering
\caption{{DOS as the indicators of topological transitions in $A_{2}$ and $A_3,A'_3$ critical point classes according to Eqs. (\ref{eq_dosa2}) and (\ref{eq_dosa3}) respectively. In both figures, red, black and blue lines denote $h=-0.2,0,$ and $+0.2$ respectively. Black lines can also be described by Eq. (\ref{eq_div}), where $A_2$ class has DOS exponent $\nu=-\frac{1}{6}$ and asymmetry parameter $\eta=\sqrt{3}$, while for $ A_3 $ class $\nu=-\frac{1}{4}$ and $\eta={\sqrt{2}}$, for $ A_3^+ $ class $\nu=-\frac{1}{4}$ and $\eta=0$.}}\label{fig_dos}
\end{figure}

As shown in Fig. \ref{fig_dos}, when $h<0$ there is a logarithmic DOS peak at $ \varepsilon=-2(|h|/3)^{3/2} $ and a DOS drop at $ \varepsilon=2(|h|/3)^{3/2} $, induced by ordinary saddle point and local energy maximum at momenta $ \pm (\sqrt{-h/3},0) $ respectively. On the contrary, when $h>0$ the DOS is a smooth function of energy without singularities. Right at $ h=0 $, DOS has a power-law divergent singularity at $E=0$ with exponent $ \nu =-\frac{1}{6} $ and particle-hole asymmetry $ \eta=\sqrt{3} $. These results are consistent with our previous expectation.

At mirror-invariant points, $A_3$ or $A'_3$ critical point can be stably realized by tuning one parameter $ h $
\begin{eqnarray}\label{eq_a3}
E =-I hp_{x}^2+p_{x}^4+I p_{y}^2,
\end{eqnarray}
where $ I=\pm $ denotes the topological index for $ A'_3 (+)$ or $A_3(-)$.
Under perturbation $h$, there will be at most 3 critical points according to the multiplicity. Since the topological index is nonzero $I\neq 0$, there will be at least one critical point in above dispersion.

The DOS of this dispersion can also be expressed in terms of K$(z)$ as \cite{magic}
\begin{equation}\label{eq_dosa3}
\begin{aligned}
\rho(\varepsilon)=\frac{{\rm sgn}(h)}{\sqrt{2}\pi^2}{\rm Re}\left\{\frac{1}{\sqrt{z_{-}}}{\rm K}\left(1-\frac{z_{+}}{z_{-}}\right)\right.\\
\left.-\Theta(I h)\frac{2i}{\sqrt{z_{+}}}{\rm K}\left(\frac{z_{-}}{z_{+}}\right)\right\}
\end{aligned}
\end{equation}
where $ z_{\pm}\equiv h\pm\sqrt{h^2+4\varepsilon} $. Here the sign function is defined as sgn$(h)=-1$ for $h<0$ and sgn$(h)=1$ for $h\geqslant 0$, and $ \Theta(h)\equiv\frac{1}{2}[1-{\rm sgn}(-h)] $ is the step function.

As shown in Fig. \ref{fig_dos}, when $h<0$ there is a logarithmic DOS peak at $ \varepsilon=-h^2/4 $ and a DOS drop at $ \varepsilon=0 $, induced by two ordinary saddle points at opposite momenta $ \pm (\sqrt{-h/2},0) $ and one local energy maximum at $\bm p=\bm 0$ respectively. On the contrary, when $h>0$ the DOS has only one logarithmic peak due to the ordinary saddle point at energy $ \varepsilon=0 $ and momentum $\bm p=\bm 0$. Right at $ h=0 $, DOS has a power-law divergent singularity at $E=0$ with exponent $ \nu =-\frac{1}{4} $ and particle-hole asymmetry $ \eta=\sqrt{2} $.

Saddle point $A_3$ and topological transition can be realized in twisted bilayer graphene by changing twist angle \cite{magic,Kerelsky}, and also related to overdoped copper oxides (Bi,Pb)$_2$Sr$_2$CuO$_{2+\delta}$ \cite{Koshicu}.

While for $A'_3$ local minimum, when $h>0$ there is a DOS drop at $ \varepsilon=-h^2/4 $ and a logarithmic DOS peak at $ \varepsilon=0 $, induced by two local energy minimum at opposite momenta $ \pm (\sqrt{-h/2},0) $ and one ordinary saddle point at $\bm p=\bm 0$ respectively. When $h<0$ the DOS has only one drop due to the ordinary energy minimum at energy $ \varepsilon=0 $ and momentum $\bm p=\bm 0$. Right at $ h=0 $, DOS has a power-law divergent singularity at $E=0$ with exponent $ \nu =-\frac{1}{6} $ and particle-hole asymmetry $ \eta=0 $.

\begin{figure}
\centering
%\subfloat[Topological transition of $A_2$ critical point.]{{\includegraphics[width=0.5\textwidth]{A2} }}
%\qquad
%\subfloat[Topological transition of $A_3$ critical point.]{{\includegraphics[width=0.5\textwidth]{A3} }}
%\qquad
\subfloat[Topological transition of $C_3\cong D_4$ critical point.]{{\includegraphics[width=0.5\textwidth]{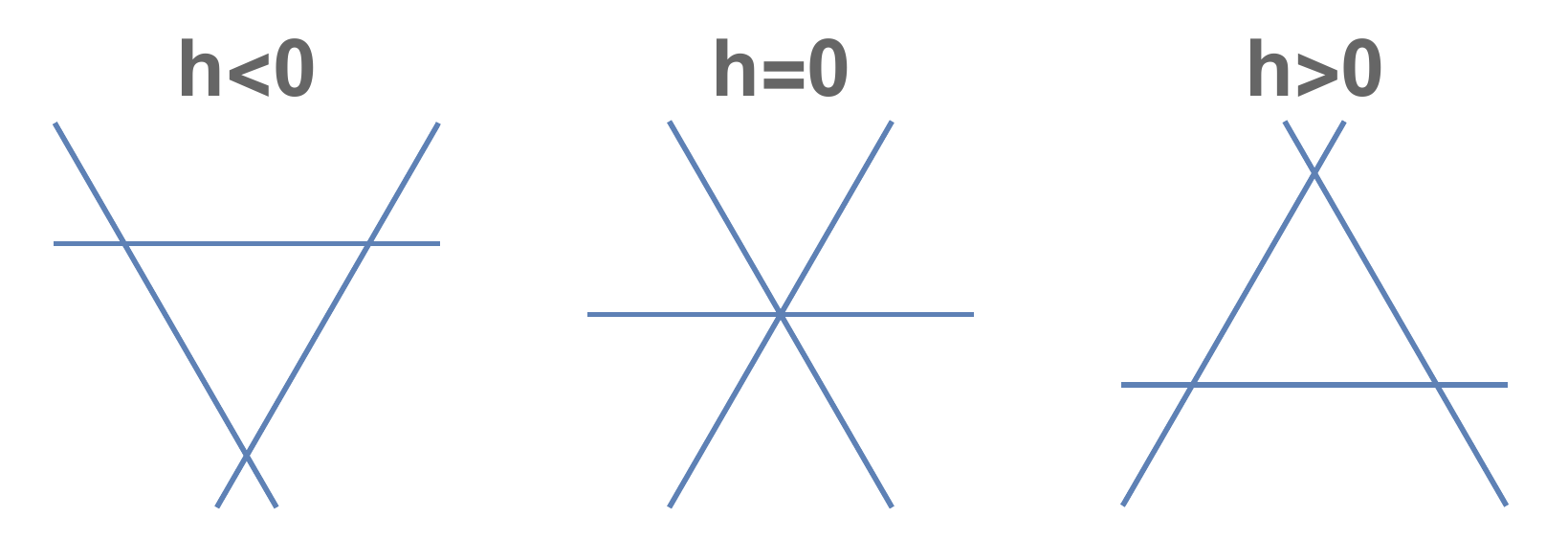} }}
\qquad
\subfloat[Topological transition of $C_4$ critical point with $r=-0.4$.]{{\includegraphics[width=0.5\textwidth]{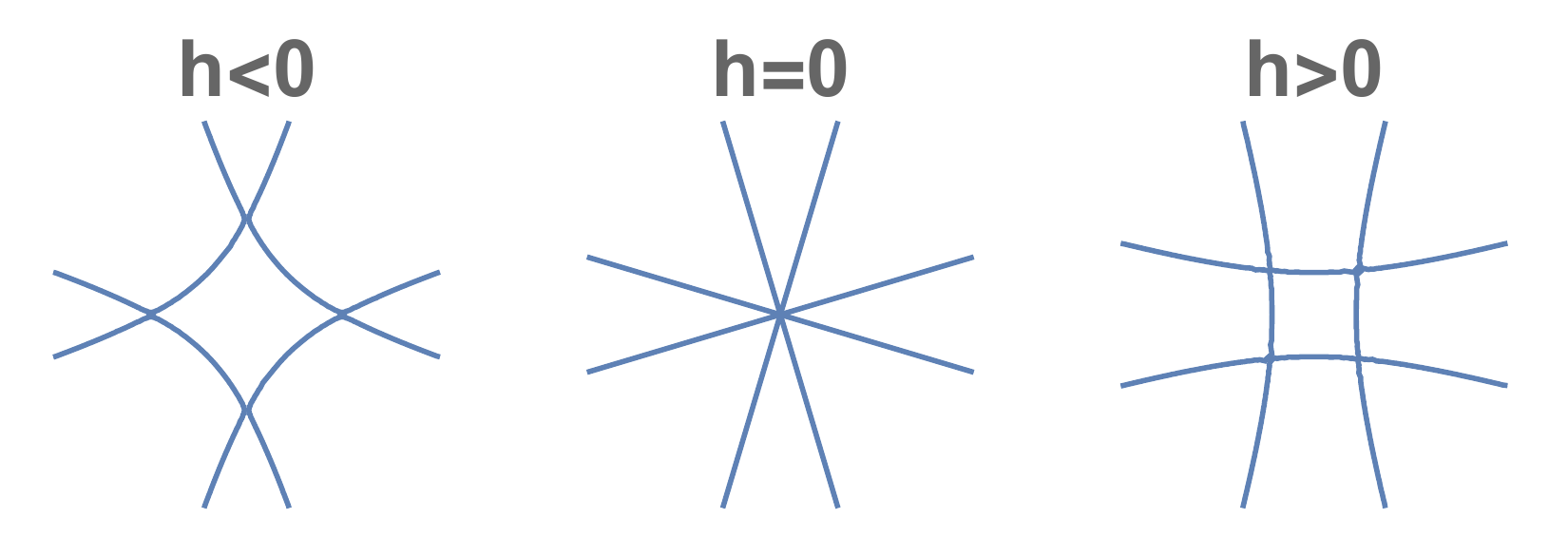} }}
\caption{Topological transitions of critical points $C_3$ and $C_4$ under single perturbations. Contours denote energy contours with saddle points.}
\label{fig_trans}
\end{figure}

With in-plane threefold rotation symmetry, $C_3$ critical point can be stably realized by tuning one parameter $h$
\begin{eqnarray}\label{eq_c3}
E=hp^2+3p_{x}^2p_{y}-p_{y}^3.
\end{eqnarray}
When $ h\neq 0 $, besides a local maximum ($h<0$) or minimum ($h>0$) at $\bm p=\bm 0$ and zero energy, there will be three ordinary saddle points at momenta related by ${\rm C}_3$ symmetry and the same energy $ 4h^3/27 $. The sign of $h$ determines the orientations of these three ordinary saddle points as shown in Fig. \ref{fig_trans} (a).
This type of critical point and topological transition can be realized in trilayer graphene on boron nitride by changing vertical electric field \cite{magic}.

With in-plane fourfold rotation symmetry, $C_4$ critical point can also be stably realized by tuning one parameter $h$
\begin{eqnarray}\label{eq_c4}
E=hp^2+{\rm Re}[(p_{x}+ip_{y})^4]+rp^4.
\end{eqnarray}
When $|r|<1$ and $ h\neq 0 $, besides a local maximum ($h<0$) or minimum ($h>0$) is always found at $\bm p=\bm 0$ and zero energy, there will be four ordinary saddle points at momentum magnitude $ p=\frac{1}{2}|h(r-{\rm sgn}h)^{-1}| $ and energy $-\frac{1}{4}h^2(r-{\rm sgn}h)^{-1}$. As shown in Fig. \ref{fig_trans} (b), the four ordinary saddle points are along the directions $ p_xp_y=0 $ if $h<0$, otherwise $p_x^2 -p_y^2=0$ if $h<0$.
This type of critical point and topological transition can be realized in Sr$_3$Ru$_2$O$_7$ by changing magnetization of Ru atoms \cite{Sr2Ru3O7,Sr2Ru3O7a}.

When $|r|>1$ and $ h\neq 0 $, a local maximum ($h<0$) or minimum ($h>0$) is always found at $\bm p=\bm 0$ and zero energy. If $hr>0$, there are no more critical points. If $hr<0$, there will be four ordinary saddle points at momentum magnitude $ p=\frac{1}{2}|h(r-{\rm sgn}h)^{-1}| $ and energy $-\frac{1}{4}h^2(r-{\rm sgn}h)^{-1}$, and four local energy minima at momentum magnitude $ p=\frac{1}{2}|h(r+{\rm sgn}h)^{-1}| $ and energy $-\frac{1}{4}h^2(r+{\rm sgn}h)^{-1}$. When $ h<0,r>0 $, ordinary saddle points are along the direction $p_xp_y=0$ while local minima are along the direction $p_x^2 -p_y^2=0$. And when $ h>0,r<0 $, ordinary saddle points are along the direction $p_x^2 -p_y^2=0$ while local minima are along the direction $p_xp_y=0$.

Notice that in these topological transitions, the total topological index is always conserved.

\subsection{An 3D example}\label{3d}
Near a critical point with cubic point group $ O_h $, and the allowed canonical dispersion and perturbation to the fourth order reads $ \mathcal{E}=\mathcal{O}_h+H $ with
\begin{eqnarray}
\mathcal{O}_{h}=p_{x}^4+p_{y}^4+p_{z}^4+rp^4,\quad H=hp^2,
\end{eqnarray}
where $ p^2\equiv p_{x}^2+p_{y}^2+p_{z}^2 $, $r$ is the continuous parameter of the canonical dispersion and $h$ is the symmetry-allowed perturbation. When $ h=0 $ and $ r\neq -1,-\frac{1}{2},-\frac{1}{3} $, the origin $ \bm p=\bm 0 $ is an isolated critical point with scaling exponents $ a=b=c=\frac{1}{4} $. Hence the DOS exponent is $ \nu=-\frac{1}{4} $ and multiplicity is $\mu=27$.

Under perturbation $h\neq 0$, the high-order critical point will split into at most $\mu=$27 ordinary critical points, which can be grouped into four classes according to symmetry. There will be 1 ordinary extremum at origin $ \bm p=\bm 0 $ with symmetry $O_h$, 6 ordinary saddle points at three axes $p_i=p_j=0,p_{k}=\pm\sqrt{\frac{-1}{2}h/(1+r)}$ with symmetry $C_{4v}$, 12 ordinary saddle points at six in-plane diagonals $|p_i|=|p_j|=\sqrt{\frac{-1}{2}h/(1+2r)}, p_k=0$ with symmetry $C_{2v}$, and 8 ordinary minimum at four main diagonals $ |p_x|=|p_y|=|p_z|=\sqrt{\frac{-1}{2}h/(1+3r)} $ with symmetry $C_{3v}$. Here $ (i,j,k) $ is a permutation of $(x,y,z)$, and critical points within each group are also related by $O_h$ symmetry.

When $ r>-\frac{1}{3} $, the high-order critical point is a local minimum with topological index $I=1$. Under perturbation $h<0$, it will split into 27 ordinary critical points: 1 ordinary maximum at origin, 6 ordinary saddle points with $ I=+1 $ at three axes, 12 ordinary saddle points with $ I=-1 $ at six in-plane diagonals, and 8 ordinary minima at four main diagonals. %As a result, the total topological index of the split critical points is $ -1+6-12+8=+1 $. 
Under perturbation $h>0$, it does not split.

When $ -\frac{1}{2}<r<-\frac{1}{3} $, the high-order critical point is a saddle point with topological index $I=-7$. Under perturbation $h<0$, it will split into 19 ordinary critical points: 1 ordinary maximum at origin, 6 ordinary saddle points with $ I=+1 $ at three axes, and 12 ordinary saddle points with $ I=-1 $ at six in-plane diagonals. %The total topological index of the split critical points is $ -1+6-12=-7 $. 
Under perturbation $h>0$, it will split into 9 ordinary critical points: 1 ordinary minimum at origin, and 8 ordinary maxima at four main diagonals. %The total topological index of the split critical points in this case is also that of the high-order critical point $ +1-8=-7 $.

When $ -1<r<-\frac{1}{2} $, the high-order critical point is a saddle point with topological index $I=5$. Under perturbation $h<0$, it will split into 7 ordinary critical points: 1 ordinary maximum at origin, and 6 ordinary saddle points with $ I=+1 $ at three axes. %As a result, the total topological index of the split critical points is $ -1+6=+5 $. 
Under perturbation $h>0$, it will split into 21 ordinary critical points: 1 ordinary minimum at origin, 8 ordinary maxima at four main diagonals and 12 ordinary saddle points with $ I=+1 $ at six in-plane diagonals. %The total topological index of the split critical points is also $ +1-8+12=+5 $.

When $ r< -1 $, the high-order critical point is a local maximum with topological index $I=-1$. Under perturbation $h<0$, it does not split. Under perturbation $h>0$, it will split into 27 ordinary critical points: 1 ordinary minimum at origin, 6 ordinary maxima at three axes, 8 ordinary saddle points with $ I=-1 $ at four main diagonals and 12 ordinary saddle points with $ I=+1 $ at six in-plane diagonals. %The total topological index of the split critical points is that of high-order critical point $ +1-6-8+12=+1 $.

In all cases discussed above, the total topological index does not change with perturbations, and the distribution of total critical points preserve $O_h$ symmetry.

\begin{table*}
\setlength{\tabcolsep}{12pt}
\renewcommand{\arraystretch}{2}
\centering
\begin{tabular}{c|c|c|c} \hline
Parameter Range & Topology & Split under Perturbation $ h<0 $ & Split under Perturbation $ h>0 $\\%[0.2cm]
\hline\hline
$\displaystyle r>-\frac{1}{3}$&$\displaystyle n_e=1,\quad n_h=0,\quad I=+1 $
&$\displaystyle -O_{h}+6C_{4v}+8C_{3v}-12C_{2v}$ &$\displaystyle O_{h}$\\[0.2cm]
\hline
$\displaystyle -\frac{1}{2}<r<-\frac{1}{3}$&$\displaystyle n_e=1,\quad n_h=8,\quad I=-7 $
&$\displaystyle -O_{h}+6C_{4v}-12C_{2v}$ &$\displaystyle O_{h}-8C_{3v}$\\[0.2cm]
\hline
$\displaystyle -1<r<-\frac{1}{2}$&$\displaystyle n_e=6,\quad n_h=1,\quad I=+5 $
&$\displaystyle -O_{h}+6C_{4v}$ &$\displaystyle O_{h}-8C_{3v}+12C_{2v}$\\[0.2cm]
\hline
$\displaystyle r<-1$&$\displaystyle n_e=0,\quad n_h=1,\quad I=-1 $
&$\displaystyle -O_{h}$ &$\displaystyle O_{h}-6C_{4v}-8C_{3v}+12C_{2v}$\\[0.2cm]
\hline
\end{tabular}
\caption{The 3D high-order critical point under perturbation. Here $ -O_{h}+6C_{4v}+8C_{3v}-12C_{2v} $ means 1 critical point at origin (symmetry $O_h$) with topological charge $I=-1$, 6 critical points at three axes (symmetry $C_{4v}$) with topological charge $I=+1$, 8 at four main diagonals (symmetry $C_{3v}$) with $I=+1$ and 12 at six in-plane diagonals (symmetry $C_{2v}$) with $I=-1$. In all cases, the distribution of total critical points preserve $O_h$ symmetry.}
\label{T1}
\end{table*}

To summarize, based on general principles of topology, scaling and symmetry, we systematically describe and classify critical points from 1D to 3D, especially high-order critical points in 2D. Our results include ordinary VHS as a natural corollary, and could be relevant in realistic materials such as 2D moir\'e systems \cite{magic,Kerelsky}, ruthenate \cite{Sr2Ru3O7,Sr2Ru3O7a,SrRuO0,SrRuO1,SrRuO2,SrRuO3} and cuprate \cite{Koshicu,EVHS2cu,EVHS5cu,EVHS6cu,Tsuei1,Tsuei2,Castro,Piriou,Tallon} materials. 
As the DOS near high-order critical points can be power-law divergent and particle-hole asymmetric, we expect high-order critical points can play important roles in transport phenomena and interaction effects, which may help us to understand correlated phases such as unconventional superconductivity and non-Fermi liquid.

{\it Note added}: While we were completing this manuscript, we became aware
of similar work also being completed by Anirudh Chandrasekaran, Alex Shtyk, Joseph J.
Betouras, and Claudio Chamon. We point the reader to that reference as well. We thank Zhi-Cheng Yang for
recognizing that the works were similar, and bringing the two groups
into contact.

\section*{Acknowledgment}
We thank Yang Zhang, Hiroki Isobe and Zhen Bi for inspiring discussions. This work is
supported by DOE Office of Basic Energy Sciences, Division of Materials Sciences and Engineering under Award
de-sc0010526. LF is partly supported by the David and
Lucile Packard Foundation.

\appendix
\setcounter{figure}{0}
\renewcommand{\thefigure}{\arabic{figure}}
\setcounter{equation}{0}
\renewcommand{\theequation}{\arabic{equation}}

\section{Coordinate Transforms}\label{ct}
In this Appendix, we derive the canonical dispersions of simple critical points. For a given simple critical point, we first write down all terms satisfying the same scaling property to form the bare dispersion, and then cast this bare dispersion into a parameter-free binomial form by appropriate coordinate transforms.

According to scaling exponents, the bare dispersion of $A_{2n-1}$ has three terms
\begin{eqnarray}\label{eq_bA}
E=\alpha p_{x}^{2n}+\beta p_{y}^2 +\gamma p_{x}^n p_{y}.
\end{eqnarray}
Under coordinate transform
\begin{eqnarray}
\tilde{p}_{x}=p_{x},\quad\tilde{p}_y=p_{y}+\frac{\gamma}{2\beta}p_{x}^n
\end{eqnarray}
the dispersion becomes binomial
\begin{eqnarray}
E=\tilde{\alpha}\tilde{p}_{x}^{2n}+\beta\tilde{p}_{y}^2
\end{eqnarray}
with $ \tilde{\alpha}=\alpha-\gamma^2/(4\beta) $. Finally after the scaling transforms on momentum and energy
\begin{eqnarray}
\tilde{\tilde{p}}_x=\sqrt[2n]{|\tilde{\alpha}|}\tilde{p}_x,\quad \tilde{\tilde{p}}_y=\sqrt{|\beta|}\tilde{p}_y,\quad \tilde{E}={\rm sgn}(\tilde{\alpha})E
\end{eqnarray}
we arrive at the parameter-free binomial form of the canonical dispersion
\begin{eqnarray}
E={p}_{x}^{2n}+I{p}_{y}^2,
\end{eqnarray}
where we dropped the tilde accent for simplicity, and $ I\equiv {\rm sgn}(\tilde{\alpha}\beta)={\rm sgn}(4\alpha\beta -\gamma^2) $ is the topological index.

%As a result, the bare dispersion (\ref{eq_bA}) is equivalent to either $A_{2n-1}$ (\ref{eq_A}) or $A'_{2n-1}$ (\ref{eq_Ap}), and the corresponding topological index of the critical point $\bm p=\bm 0$ is $ I={\rm sgn}(4\alpha\beta -\gamma^2)=\mp $ respectively.

According to scaling exponents, the bare dispersion of $D_{2n+2}(n\geqslant 2)$ has three terms
\begin{eqnarray}\label{eq_bD}
E=\alpha p_{x}^{2}p_{y}+\beta p_{y}^{2n+1} +\gamma p_{x} p_{y}^{n+1}.
\end{eqnarray}
Under coordinate transform
\begin{eqnarray}
\tilde{p}_{x}=p_{x}+\frac{\gamma}{2\alpha}p_{y}^n,\quad\tilde{p}_y=p_{y}
\end{eqnarray}
the dispersion becomes binomial
\begin{eqnarray}
E={\alpha}\tilde{p}_{x}^{2}\tilde{p}_{y}+\tilde{\beta}\tilde{p}_{y}^{2n+1}
\end{eqnarray}
with $ \tilde{\beta}=\beta-\gamma^2/(4\alpha) $. Finally after the scaling transforms on momentum and energy
\begin{eqnarray}
\tilde{\tilde{p}}_x=\frac{\sqrt{|\alpha|}}{\beta^{\frac{1}{2n+1}}}\tilde{p}_x,\quad \tilde{\tilde{p}}_y=\beta^{\frac{1}{2n+1}}\tilde{p}_y,\quad \tilde{E}={\rm sgn}({\alpha})E
\end{eqnarray}
we arrive at the parameter-free binomial form of the canonical dispersion
\begin{eqnarray}
E={p}_{x}^{2}p_y+(I+1){p}_{y}^{2n+1},
\end{eqnarray}
where we dropped the tilde accent for simplicity, $ I\equiv {\rm sgn}(4\alpha\beta -\gamma^2)-1 $ is the topological index.

%As a result, the bare dispersion (\ref{eq_bD}) is equivalent to either $D_{2n+2}$ (\ref{eq_D}) or $D'_{2n+2}$ (\ref{eq_Dp}), and the corresponding topological index of the critical point $\bm p=\bm 0$ is $ I={\rm sgn}(4\alpha\beta -\gamma^2)=\mp $ respectively.

According to scaling exponents, the bare dispersion of $D_{4}$ has four terms
\begin{eqnarray}
E=a_1 p_{x}^{3}+a_2 p_{x}^{2}p_{y} +a_3 p_{x} p_{y}^{2}+a_4 p_{y}^3.
\end{eqnarray}
Under linear coordinate transform $ \tilde{\bm p}=A\bm p $ where $ A\in\mathbb{R}^{2\times 2} $ has four independent parameters,
this homogeneous dispersion becomes parameter-free
\begin{eqnarray}
E=\tilde{p}_{x}^{2}\tilde{p}_{y}\pm\tilde{p}_{y}^{3},
\end{eqnarray}
which is canonical dispersion $D_4$ or $D'_4$ in (\ref{eq_D}).

Except for the critical points above, bare dispersions of other simple critical points are all binomial and can be cast into parameter-free forms in Sec. \ref{sp} under scaling transforms on momentum and energy.

When the critical point has $ {\rm C}_2 $ symmetry, the homogeneous dispersion of order $N=4$ has five terms
\begin{eqnarray}
E=b_1 p_{x}^{4}+b_2 p_{x}^{3}p_{y} +b_3 p_{x}^2 p_{y}^{2}+b_4 p_{x}p_{y}^3+b_{5} p_{y}^4.
\end{eqnarray}
Under linear coordinate transform $ \tilde{\bm p}=B\bm p $ where $ B\in\mathbb{R}^{2\times 2} $ has four independent parameters,
this homogeneous dispersion can contain one parameter $r$
\begin{eqnarray}
E={\rm Re}[(\tilde{p}_{x}+i\tilde{p}_{y})^{2}]\tilde{p}^{2}+r\tilde{p}^{4},
\end{eqnarray}
which is canonical dispersion $C_2$ in (\ref{eq_C2}).

\section{Tight-binding models}\label{mod}
In this Section, we propose three tight-binding models, each with two independent tuning parameters but different symmetries, and we can realize $ A_2,A_3,A'_3, C_3,C_4,O $ critical points in these models. In all these tight-binding models, there are always at least three critical points, and the total topological index of all critical points is zero, due to constraint Eq. (\ref{eq_charge}). Except the special cases mentioned below, all other critical points of the tight-binding models are ordinary.

We first consider the following rectangular lattice tight-binding model
\begin{equation}\label{eq_mod}
H_{a}=\sum_{j}-t_{y}a^{\dagger}_{j}a_{j+\hat{\bm y}}-t_{x}a^{\dagger}_{j}a_{j+\hat{\bm x}}-t'_{x}e^{i\phi}a^{\dagger}_{j}a_{j+2\hat{\bm x}}+h.c.
\end{equation}
where $t_y,t_x,t'_x>0$ are hopping amplitudes and $ \phi $ denotes the phase of hopping. The symmetry group of this model is ${\rm C}_{1v}$ with mirror symmetry. This model could be relevant in some ruthenate systems \cite{SrRuO0,SrRuO1,SrRuO2,SrRuO3}.

\begin{figure}
\includegraphics[width=0.5\textwidth]{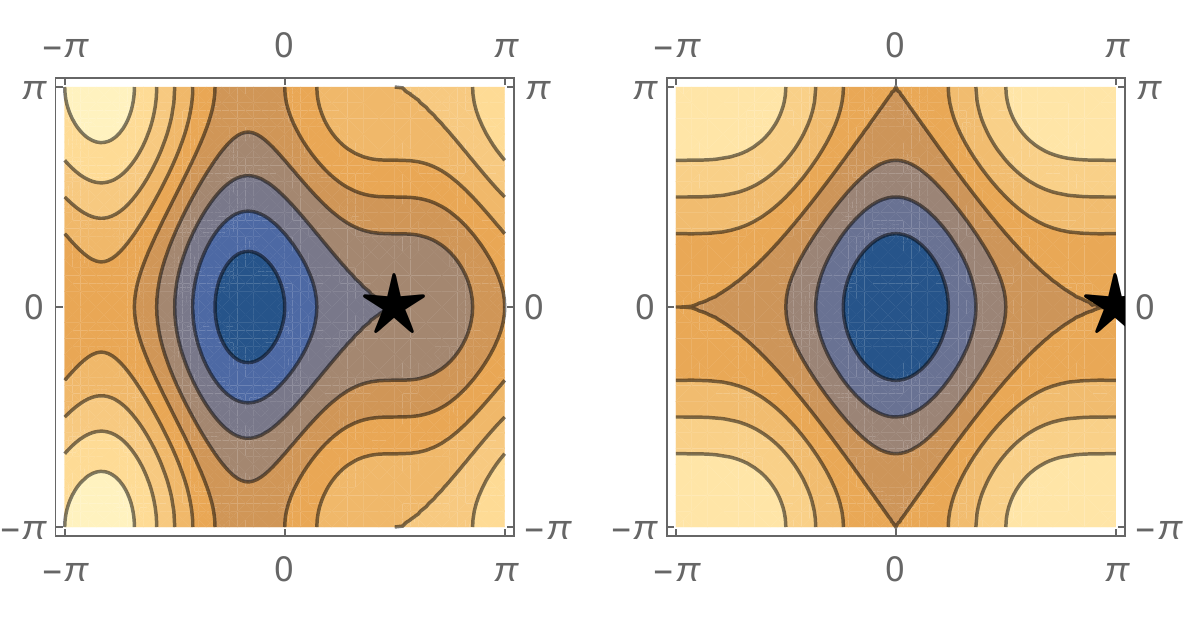}
\centering
\caption{{Energy contours of model Hamiltonian (\ref{eq_mod}) with $ t'_{x}=\frac{1}{2}t_x,\phi=\frac{1}{2}\pi $ (Left panel) and $ t'_x=\frac{1}{4}t_x,\phi=0 $ (Right panel) respectively. In the Left panel, $A_2$ class saddle point is realized at $ (\frac{1}{2}\pi,0) $, while $A_3$ is realized at $ (\pi,0) $ in the Right panel, both marked by stars. In both cases $ t_y=t_x $.}}\label{fig_mod}
\end{figure}

When $ 1/4<t'_{x}/t_{x}\leqslant 1/2 $, we can always find with an appropriate phase $\phi=\phi(t'_{x}/t_{x})$ so that a saddle point of $A_2$ class is realized at a point of $ k_{y}=0 $ line. An example is shown in the left panel of Fig. \ref{fig_mod}, where we find 4 critical points: two saddle points $ A_1,A_2 $ at $ k_{y}=0 $ line, one local maximum and one local minimum which are both ordinary (i.e. $ A'_{1} $).

When $ t'_{x}/t_{x}=1/4 $, if $ \phi =0 $ a saddle point $A_3$ is realized at $\bm X=(\pi,0) $, and a local maximum $A'_3$ is realized at $ \bm M =(\pi,\pi) $. If $ \phi =\pi $, a saddle point $A_3$ is realized at $\bm Y=(0,\pi) $, and a local minimum $A'_3$ is realized at $ \bm\Gamma =(0,0) $. In both cases, the system has emergent C$_{2v}$ symmetry. In the example shown in the right panel of Fig. \ref{fig_mod}, where we find 4 critical points: one saddle point $ A_13 $ at $ \bm X $, one saddle point $A_1$ at $ \bm Y $, one local maximum $A'_3$ at $\bm M$ and one local minimum $ A'_{1} $ at $\bm\Gamma$.

The tight-binding model (\ref{eq_mod}) is invariant under the operation $ \phi\to\phi+\pi,\bm k\to\bm k+\bm M,E\to -E $. Thus $ \phi =0 $ and $ \phi =\pi $ are related as discussed above, and $ \phi\in [0,\pi) $ is sufficient to capture the $\phi$-dependence of this system.

Next we consider a square lattice tight-binding model with nearest, next-nearest and next-next-nearest neighbor hopping terms
\begin{equation}\label{eq_mod1}
H_{b}=-\sum_{n=1}^3t_{n}\sum_{\langle ij\rangle_{n}}b^{\dagger}_{i}b_{j},
\end{equation}
where $t_1,t_2,t_3\in\mathbb{R}$ are hopping integrals. The symmetry group of this model is ${\rm C}_{4v}$, which contains mirror and fourfold rotation symmetries. This model can be relevant in some cuprate systems \cite{Koshicu,EVHS2cu,EVHS5cu,EVHS6cu}.

\begin{figure}
\includegraphics[width=0.5\textwidth]{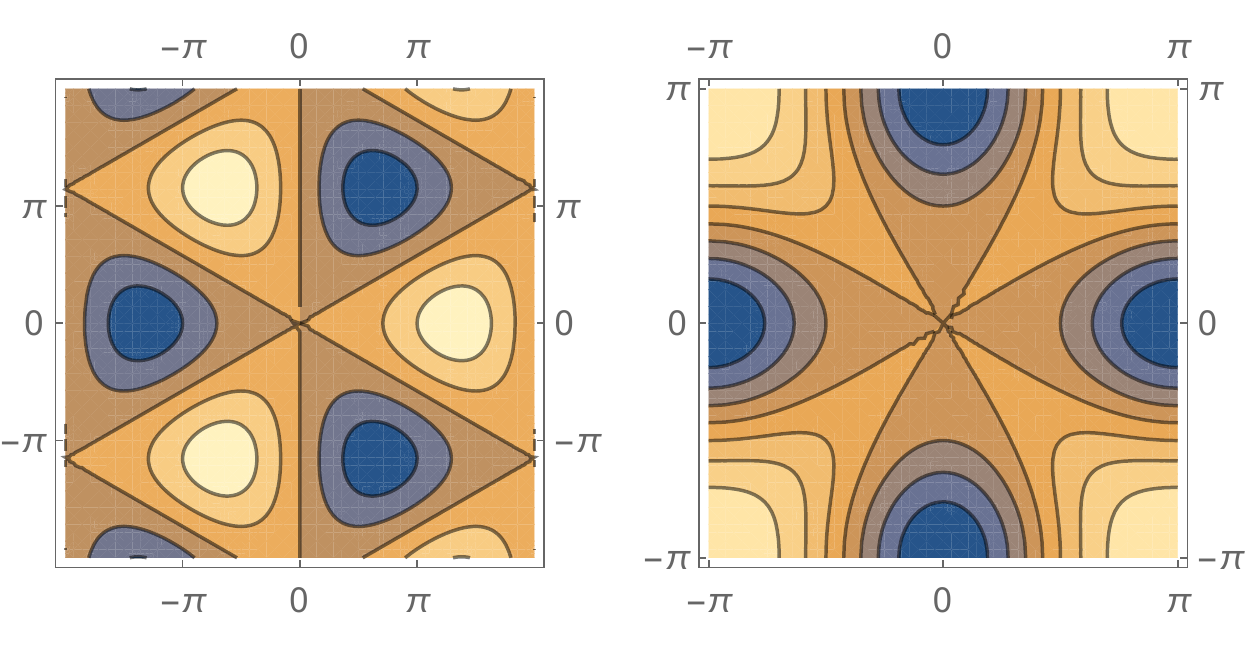}
\centering
\caption{{Left panel: Energy contours of model Hamiltonian (\ref{eq_mod2}) with $ t_{2}=0,\Phi=\frac{1}{2}\pi $ and $C_3$ class saddle point is realized at $\Gamma$ point. Right panel: Energy contours of model Hamiltonian (\ref{eq_mod1}) with $ t_{2}=-1.5t_1,t_3=0.5t_1 $ and $C_4$ class saddle point is realized at $ \Gamma $ point.}}\label{fig_modc}
\end{figure}

When $ \frac{1}{2}t_1+t_2+2t_3=0 $, critical point $C_4$ is realized at $\Gamma$ point (Fig. \ref{fig_modc} Right panel),
%with continuous parameter $ r=3(t_1+6t_2+16t_3)/(t_1-4t_2+16t_3) $.
where $t_3>0$ corresponds to a saddle point while $t_3<0$ a local extremum.
When $ -\frac{1}{2}t_1+t_2+2t_3=0 $, critical point $C_4$ is realized at $\bm M=(\pi,\pi) $ point,
%with continuous parameter $ r=3(t_1-6t_2-16t_3)/(t_1+4t_2-16t_3) $. H
where $t_3<0$ corresponds to a saddle point while $t_3>0$ a local extremum.
In above cases, when $t_3=0$ there will be two critical lines instead of a single critical point.

When $ \pm\frac{1}{2}t_1-t_2+2t_3=0 $, critical point $A_3$ or $A'_3$ is realized at both $\bm X=(\pi,0)$ and $ \bm Y=(0,\pi) $ points due to fourfold rotation, and the topological index depends on the details.

To realize $C_3$ and $O$ class critical points, we can use a triangular lattice tight-binding model with nearest and next-nearest neighbor hopping terms \cite{magic,Constantine,TLG3}
\begin{eqnarray}\label{eq_mod2}
H_{c}=\sum_{\tau=\pm}\left\{\sum_{\langle ij\rangle}t_{1}e^{i\Phi\tau}c_{i\tau}^{\dagger}c_{j\tau}+\sum_{\langle\langle ij\rangle\rangle}t_{2}c_{i\tau}^{\dagger}c_{j\tau}\right\}.
\end{eqnarray}
where $\tau=\pm$ denotes two orbitals, $t_{1,2}>0$ are hopping amplitudes and $ \Phi $ denotes the phase of hopping.
The symmetry group of this model is ${\rm C}_{3v}$, which contains mirror and threefold rotation symmetries. This model may be relevant in moir\'e systems of trilayer graphene on boron nitride \cite{TLG1,TLG2,TLG3} and transition metal dichalcogenides \cite{Strain,Constantine}.

When $ 3t_{2}=-t_{1}\cos\Phi $ but $ \sin\Phi\neq 0 $, there will be $C_3$ class saddle point realized at $ \Gamma $ point of the BZ, as shown in the left panel of Fig. \ref{fig_modc}. In this case we have in total 3 critical points in the entire BZ: one saddle point $ C_2 $ at $ \Gamma $ point, ordinary local maximum and local minimum at $ \pm\bm K $ respectively. Here 3 is the minimum number of critical points allowed by topology (\ref{eq_charge}).

When $ 3t_{2}=-t_{1}\cos\Phi $ and $ \sin\Phi= 0 $, $ \Gamma $ point is a local extremum of $O$ class and the system has emergent ${\rm C}_{6v}$ symmetry.

The dispersion of (\ref{eq_mod2}) is invariant under transform $ \Phi\to\Phi +2\pi/3,\bm k\to\bm k+\bm K $ where $ \bm K=(4\pi/3,0) $. After this transform, the criteria of high-order critical point at $\Gamma$ point can be applied to $\pm\bm K$ points.

%In all tight-binding models of this Section, the local energy dispersion near a critical point can always be described by the corresponding canonical dispersion together with some irrelevant perturbations. As we discussed previously, relevant perturbations will generally split the critical point and eventually to ordinary ones. In the following Section we will discuss effects of relevant perturbations in detail for six specific examples.

\end{document}